# On the short time prediction of earthquakes in the Balkan- Black Sea region based on geomagnetic field measurements and tide gravitational potential behavior

Institute for Nuclear Research and Nuclear Energy, Bulgarian Academy of Sciences,
Strachimir Chterev Mavrodiev
Sofia, August 2002
E-mail: Mavrodi@inrne.bas.bg

**Abstract**: The paper is a first attempt for statistical estimation of method for a short time prediction of incoming earthquakes in the Balkans and Black Sea region from January to July, 2002. The essence of the discovery is that the geomagnetic local "quake" is a precursor of future earthquake, which time is determined by the tide gravitational potential behavior. A Balkans Black Sea region Earthquake's "When, where" Prediction Network is proposed.

## Introduction

The prediction of the earthquakes' time (date, hour, minute), epicenter (coordinates, depth), Magnitude (Richter scale), the destructive power on the surface and the danger from Tsunami and Volcano eruption, is one long time non-solved science problem. Moreover, in the last years many scientists and International Science Institutions have arrived to the conclusion that "Although a great deal is known about *where* earthquakes are likely, there is currently no reliable way to predict the days or months *when* an event will occur in any specific location" [1].

The explanation is because of the very high complexity of the Earth crust dynamics and very "scare" time and space points of observation, the non-adequate mathematical methods of analysis, it is not possible to obtain an unique signal from the data [2]. Another reason is that, in difference with elementary particle physics, in the area of Earth sciences there is not a unified data bank with the same formats, theory, formulae and solutions. Nevertheless, many scientists and Institutions continue their efforts looking for the possible solutions of the earthquake prediction problem [3].

Our methodology is based on the impressive development of the 20-th Century science. Here we have to mention the physicist, chemist and biologist N. I. Vernadsky [4], the theoretical physicist F. Dayson [5] and of the biologist N. M. Sissakian [6]. Their scientific work points out the birth of the environmental science, whose subject is the whole Earth's ecosystem with its main component - the human civilization. Some people call this science "sustainable development". The author prefers the more exact term "harmonic existence".

## Measurements and analysis

The geomagnetic field (GMF) projection is measured with a detector (know- how of JINR, Dubna, Boris Vasiliev) with absolute precision less than or equal to 1 nano Tesla [nT] at a sampling rate of 2.4 samples per second. Each minute the mean value $GMF_m$ is given by

$$GMF_m = \sum_{i=1,Nm} GMF_i / Nm,$$

the mean value of the error $\Delta GMF_m$ is

$$\Delta GMF_m = \sum_{i=1,Nm} \Delta GMF_i / Nm,$$

the mean square deviation $\sigma GMF_m$ is

$$\sigma GMF_m = (sqrt(\sum_{i=1,Nm} (GMF_i - GMFm)^2)/ Nm ,$$

and the mean square deviation of the error $\sigma\Delta GMF_m$ is

$$\sigma\Delta GMF_m = (sqrt(\sum_{i=1,Nm} (\Delta GMF_i - \Delta GMF_m)^2)/ Nm.$$

So, for 24 hours $N_d = 1440$ data: the time (ddmmyy hhmm) and quartet $GMF_m$, $\Delta GMF_m$ , $\sigma GMF_m$ $\sigma\Delta GMF_m$ of data are recorded.

Figure 1 illustrates the behavior of $GMF_m$ (upper picture) and $\sigma GMF_m$, $\sigma\Delta GMF_m$ (down picture) in a day without a Signal for near future event.

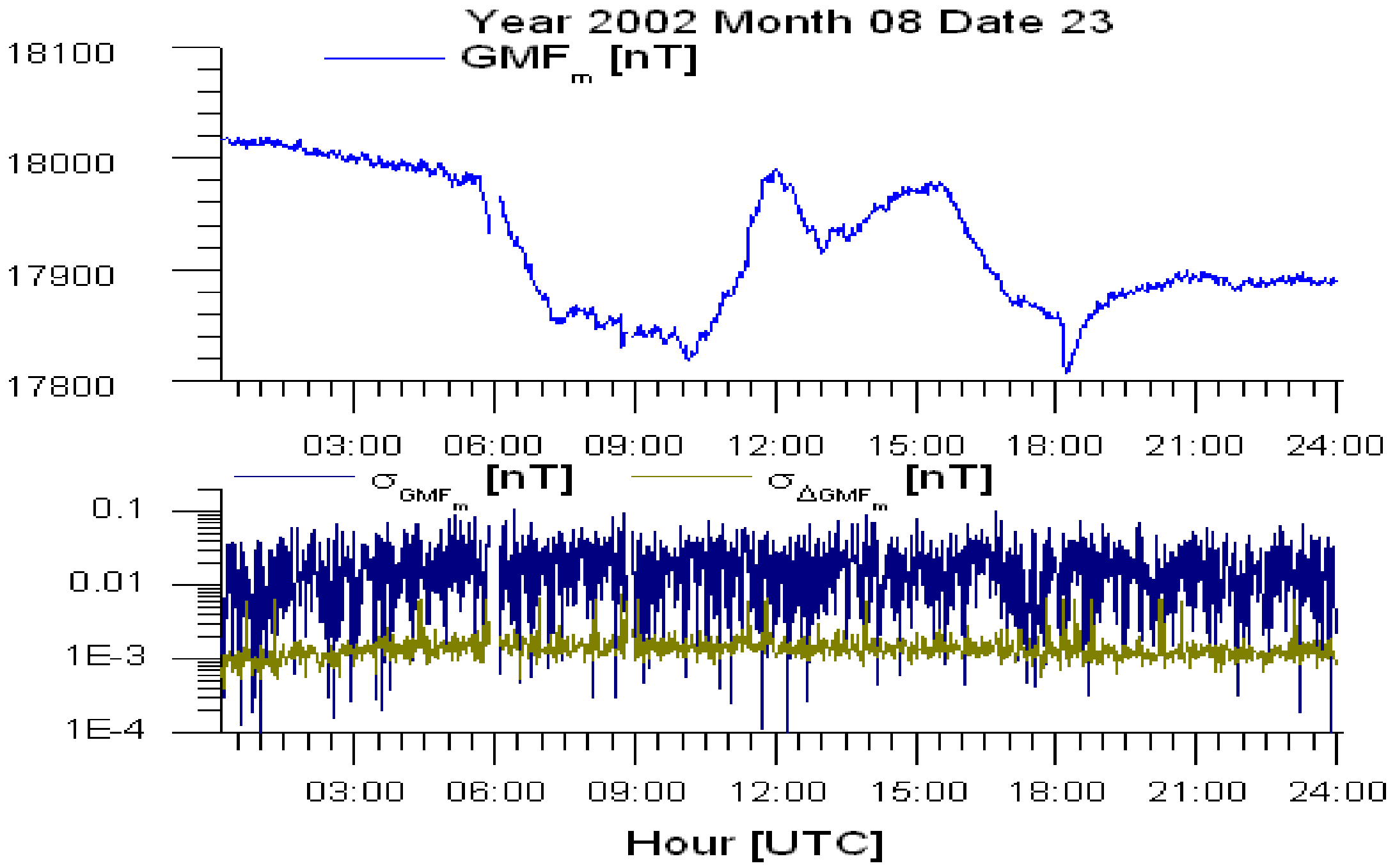

Figure 1.

Figure 2 illustrates the behavior of $GMF_m$ (upper picture) and $\sigma GMF_m$, $\sigma\Delta GMF_m$ (down picture) in a day with a Signal for near future event.

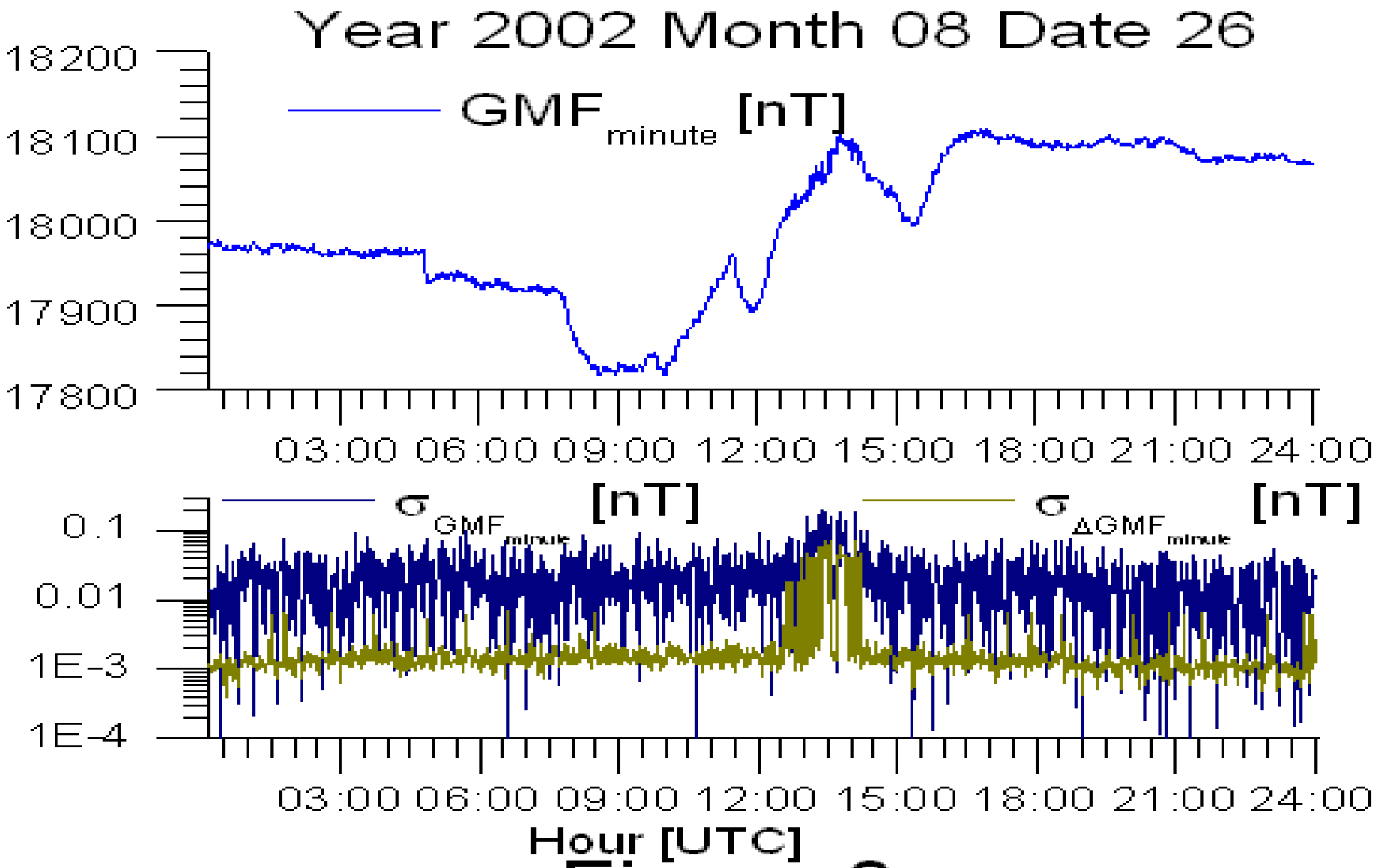

Figure 2

The every day data are collected in a month collected data- base. The upper picture of the next Figure 3 presents the $GMF_m$ behavior for every minute of the month and the down picture presents the behavior of a signal Sig with error $\Delta Sig$, which are averaged for the day values of $\sigma GMF_m$ and $\sigma\Delta GMF_m$ :

$$Sig = \sum_{i=1,Nd} \sigma GMF_m / Nd \quad \text{and} \quad \Delta Sig = \sum_{i=1,Nd} \sigma\Delta GMF_m / Nd$$

**EQTable2002** **Jan**

| ddmmyy hhmm | Lat | Lon | kmSof | Dep | Mag | SChtM | ddmmyy hhmm | Lat | Lon | kmSof | Dep | Mag | SChtM |
|---|---|---|---|---|---|---|---|---|---|---|---|---|---|
| 01/01/2002 15:08 | 37.94 | 21.90 | 516 | 5 | 3.3 | 12.4 | 20/01/2002 13:38 | 38.62 | 23.79 | 424 | 5 | 3.7 | 20.6 |
| 01/01/2002 22:15 | 37.34 | 21.76 | 585 | 5 | 4.3 | 12.6 | 21/01/2002 00:29 | 35.54 | 23.33 | 762 | 5 | 3.7 | 6.4 |
| 02/01/2002 10:10 | 37.30 | 22.08 | 580 | 3 | 3.2 | 9.5 | 21/01/2002 00:48 | 38.91 | 22.66 | 392 | 5 | 3.1 | 20.1 |
| 02/01/2002 11:11 | 38.41 | 25.85 | 530 | 39 | 3.4 | 12.1 | 21/01/2002 14:34 | 38.66 | 27.88 | 663 | 10 | 4.8 | 10.9 |
| 03/01/2002 00:17 | 39.68 | 23.03 | 303 | 3 | 3.5 | 38.2 | 21/01/2002 16:25 | 38.39 | 21.73 | 475 | 4 | 3.2 | 14.2 |
| 04/01/2002 02:08 | 35.02 | 26.25 | 886 | 4 | 3.9 | 5.0 | 21/01/2002 19:36 | 38.38 | 21.82 | 473 | 5 | 3.1 | 13.9 |
| 04/01/2002 02:32 | 38.38 | 21.76 | 475 | 5 | 3.3 | 14.6 | 21/01/2002 23:09 | 36.26 | 22.21 | 691 | 5 | 3.7 | 7.8 |
| 04/01/2002 16:20 | 37.28 | 21.92 | 586 | 5 | 3.3 | 9.6 | 21/01/2002 23:59 | 39.03 | 24.35 | 395 | 21 | 2.9 | 18.6 |
| 05/01/2002 03:35 | 38.86 | 21.99 | 416 | 5 | 3.3 | 19.1 | 22/01/2002 01:05 | 40.58 | 20.15 | 396 | 10 | 3.3 | 21.0 |
| 06/01/2002 06:09 | 37.73 | 22.27 | 529 | 33 | 4.1 | 14.7 | 22/01/2002 04:37 | 39.31 | 21.49 | 393 | 5 | 2.9 | 18.8 |
| 06/01/2002 12:28 | 38.32 | 22.53 | 459 | 37 | 2.9 | 13.8 | 22/01/2002 04:53 | 35.79 | 26.62 | 825 | 88 | 6.3 | 9.3 |
| 06/01/2002 15:36 | 39.38 | 23.74 | 340 | 5 | 4.1 | 35.4 | 22/01/2002 17:37 | 39.40 | 25.78 | 438 | 36 | 3.4 | 17.7 |
| 07/01/2002 11:09 | 34.87 | 24.17 | 843 | 48 | 4.1 | 5.8 | 22/01/2002 20:41 | 40.63 | 20.17 | 391 | 5 | 3.0 | 19.6 |
| 07/01/2002 16:57 | 38.59 | 23.62 | 425 | 5 | 3.1 | 17.1 | 22/01/2002 22:02 | 40.40 | 23.63 | 227 | 3 | 2.9 | 56.5 |
| 07/01/2002 18:41 | 37.39 | 22.09 | 570 | 23 | 2.8 | 8.6 | 23/01/2002 12:53 | 38.46 | 25.54 | 508 | 7 | 3.2 | 12.4 |
| 08/01/2002 00:57 | 38.95 | 24.45 | 407 | 33 | 3.1 | 18.8 | 23/01/2002 15:12 | 39.99 | 19.54 | 488 | 5 | 3.1 | 13.0 |
| 08/01/2002 14:17 | 36.25 | 30.03 | 1019 | 33 | 3.9 | 3.8 | 23/01/2002 15:18 | 39.99 | 19.47 | 495 | 14 | 3.2 | 13.1 |
| 08/01/2002 17:13 | 39.20 | 23.49 | 357 | 19 | 3.3 | 26.0 | 23/01/2002 18:24 | 34.60 | 24.83 | 884 | 5 | 3.8 | 4.9 |
| 09/01/2002 09:35 | 37.83 | 21.25 | 552 | 33 | 4.4 | 14.4 | 23/01/2002 20:10 | 38.55 | 27.71 | 657 | 31 | 3.9 | 9.1 |
| 09/01/2002 14:11 | 36.10 | 22.83 | 701 | 33 | 4.4 | 9.0 | 23/01/2002 21:22 | 38.51 | 27.62 | 652 | 0 | 4.1 | 9.7 |
| 09/01/2002 16:13 | 39.12 | 24.12 | 378 | 37 | 3.0 | 21.1 | 24/01/2002 02:20 | 35.94 | 27.50 | 860 | 59 | 3.8 | 5.1 |
| 10/01/2002 01:06 | 40.05 | 19.66 | 474 | 5 | 3.1 | 13.8 | 24/01/2002 03:01 | 41.14 | 20.02 | 382 | 9 | 2.6 | 17.8 |
| 10/01/2002 02:42 | 40.04 | 19.59 | 481 | 5 | 3.0 | 13.0 | 24/01/2002 03:59 | 38.07 | 22.82 | 483 | 5 | 3.6 | 15.5 |
| 11/01/2002 02:22 | 40.04 | 20.64 | 388 | 34 | 3.1 | 20.6 | 25/01/2002 14:38 | 38.50 | 25.75 | 516 | 31 | 3.5 | 13.1 |
| 11/01/2002 03:15 | 39.53 | 20.40 | 447 | 5 | 3.1 | 15.5 | 25/01/2002 16:09 | 38.84 | 20.49 | 498 | 6 | 2.9 | 11.7 |
| 11/01/2002 08:12 | 35.53 | 23.67 | 764 | 20 | 3.6 | 6.2 | 25/01/2002 17:29 | 38.56 | 25.66 | 505 | 34 | 3.5 | 13.7 |
| 11/01/2002 17:29 | 40.12 | 21.31 | 330 | 5 | 3.2 | 29.4 | 25/01/2002 17:54 | 38.18 | 20.63 | 550 | 10 | 4.5 | 14.9 |
| 11/01/2002 18:44 | 39.97 | 21.39 | 338 | 5 | 3.1 | 27.2 | 26/01/2002 07:47 | 34.86 | 32.65 | 1340 | 10 | 3.1 | 1.7 |
| 11/01/2002 19:39 | 40.17 | 21.13 | 339 | 6 | 3.0 | 26.1 | 26/01/2002 07:51 | 34.86 | 32.65 | 1340 | 10 | 3.2 | 1.8 |
| 12/01/2002 04:21 | 37.20 | 21.71 | 601 | 5 | 2.7 | 7.5 | 26/01/2002 10:54 | 38.77 | 21.55 | 444 | 5 | 3.3 | 16.8 |
| 12/01/2002 09:30 | 35.14 | 22.99 | 806 | 74 | 3.9 | 6.0 | 26/01/2002 11:03 | 37.17 | 21.00 | 631 | 10 | 4.2 | 10.6 |
| 12/01/2002 20:40 | 38.29 | 21.94 | 478 | 5 | 2.8 | 12.3 | 26/01/2002 11:07 | 37.17 | 21.15 | 624 | 5 | 3.7 | 9.5 |
| 12/01/2002 21:42 | 38.38 | 21.70 | 477 | 8 | 2.8 | 12.3 | 26/01/2002 11:24 | 37.17 | 21.18 | 623 | 5 | 3.5 | 9.0 |
| 12/01/2002 22:48 | 37.68 | 21.31 | 565 | 5 | 2.9 | 9.1 | 26/01/2002 12:22 | 37.06 | 20.93 | 645 | 10 | 4.1 | 9.9 |
| 12/01/2002 23:16 | 39.10 | 24.27 | 384 | 33 | 3.1 | 21.0 | 26/01/2002 13:18 | 37.07 | 20.99 | 641 | 5 | 3.5 | 8.5 |
| 13/01/2002 07:31 | 38.50 | 20.64 | 519 | 5 | 3.2 | 11.9 | 26/01/2002 13:25 | 37.19 | 21.16 | 622 | 5 | 3.6 | 9.3 |
| 13/01/2002 07:50 | 38.59 | 24.78 | 457 | 2 | 3.4 | 16.3 | 26/01/2002 14:57 | 37.04 | 20.86 | 650 | 5 | 3.8 | 9.0 |
| 13/01/2002 08:19 | 38.66 | 23.84 | 421 | 10 | 2.9 | 16.4 | 26/01/2002 15:31 | 37.06 | 20.90 | 646 | 5 | 3.7 | 8.9 |
| 13/01/2002 08:46 | 38.50 | 25.53 | 503 | 5 | 3.9 | 15.4 | 26/01/2002 20:05 | 37.11 | 21.07 | 634 | 10 | 4.7 | 11.7 |
| 14/01/2002 11:16 | 39.93 | 21.44 | 338 | 9 | 2.8 | 24.5 | 26/01/2002 20:15 | 37.12 | 20.98 | 637 | 5 | 3.9 | 9.6 |
| 14/01/2002 13:46 | 38.65 | 23.64 | 419 | 5 | 3.5 | 20.0 | 26/01/2002 20:21 | 37.11 | 21.01 | 636 | 5 | 3.9 | 9.6 |
| 14/01/2002 23:22 | 39.04 | 21.91 | 400 | 5 | 3.0 | 18.7 | 26/01/2002 20:24 | 37.24 | 21.09 | 620 | 10 | 4.4 | 11.5 |
| 15/01/2002 12:14 | 38.87 | 25.73 | 481 | 30 | 3.5 | 15.1 | 26/01/2002 20:45 | 37.04 | 20.98 | 645 | 5 | 3.4 | 8.2 |
| 15/01/2002 13:53 | 40.26 | 19.66 | 461 | 5 | 3.2 | 15.1 | 26/01/2002 21:31 | 37.11 | 20.98 | 638 | 5 | 3.7 | 9.1 |
| 15/01/2002 21:56 | 38.60 | 26.69 | 571 | 26 | 3.5 | 10.7 | 26/01/2002 22:05 | 37.07 | 20.92 | 644 | 5 | 3.3 | 8.0 |
| 16/01/2002 00:58 | 34.90 | 32.67 | 1339 | 25 | 2.6 | 1.5 | 26/01/2002 22:18 | 37.13 | 20.76 | 646 | 5 | 3.4 | 8.2 |
| 16/01/2002 09:12 | 34.52 | 31.97 | 1307 | 25 | 2.9 | 1.7 | 26/01/2002 22:43 | 37.17 | 21.13 | 625 | 5 | 3.3 | 8.4 |
| 16/01/2002 18:24 | 41.31 | 19.83 | 395 | 56 | 4.0 | 25.6 | 27/01/2002 01:07 | 37.09 | 20.97 | 640 | 5 | 3.3 | 8.1 |
| 16/01/2002 18:30 | 34.95 | 23.38 | 827 | 57 | 3.9 | 5.7 | 27/01/2002 02:08 | 37.15 | 21.04 | 631 | 5 | 3.4 | 8.5 |
| 17/01/2002 02:08 | 37.71 | 21.10 | 571 | 5 | 3.5 | 10.7 | 27/01/2002 03:48 | 36.92 | 20.96 | 658 | 5 | 3.5 | 8.1 |
| 17/01/2002 02:48 | 37.50 | 31.40 | 1058 | 33 | 4.2 | 3.8 | 27/01/2002 04:54 | 39.23 | 24.21 | 369 | 12 | 3.4 | 25.0 |
| 17/01/2002 21:32 | 38.63 | 25.83 | 509 | 39 | 3.4 | 13.1 | 27/01/2002 06:08 | 37.65 | 20.52 | 607 | 9 | 3.1 | 8.4 |
| 17/01/2002 21:57 | 39.03 | 23.97 | 383 | 5 | 3.6 | 24.5 | 27/01/2002 08:44 | 38.36 | 22.30 | 460 | 19 | 4.2 | 19.9 |
| 17/01/2002 23:17 | 34.51 | 32.01 | 1311 | 25 | 3.0 | 1.8 | 27/01/2002 12:48 | 37.12 | 21.03 | 635 | 5 | 3.0 | 7.5 |
| 18/01/2002 02:02 | 39.11 | 24.24 | 382 | 7 | 3.0 | 20.5 | 27/01/2002 16:57 | 39.18 | 24.19 | 373 | 30 | 3.3 | 23.7 |
| 18/01/2002 05:17 | 35.54 | 26.48 | 843 | 37 | 3.6 | 5.1 | 27/01/2002 17:02 | 38.49 | 25.56 | 506 | 24 | 3.4 | 13.3 |
| 18/01/2002 11:53 | 38.17 | 20.60 | 552 | 5 | 3.2 | 10.5 | 27/01/2002 20:12 | 37.08 | 21.02 | 639 | 5 | 3.0 | 7.4 |
| 18/01/2002 12:52 | 39.17 | 24.25 | 376 | 31 | 3.6 | 25.4 | 27/01/2002 20:47 | 38.56 | 25.72 | 509 | 38 | 3.3 | 12.8 |
| 18/01/2002 14:31 | 37.42 | 42.74 | 2236 | 33 | 4.4 | 0.9 | 28/01/2002 02:25 | 38.19 | 23.81 | 472 | 5 | 2.9 | 13.0 |
| 19/01/2002 00:04 | 38.33 | 23.89 | 458 | 14 | 3.1 | 14.8 | 28/01/2002 04:15 | 39.99 | 21.70 | 316 | 5 | 3.3 | 33.0 |
| 19/01/2002 00:07 | 36.85 | 21.76 | 637 | 5 | 3.2 | 7.9 | 28/01/2002 11:03 | 37.12 | 21.01 | 635 | 5 | 3.1 | 7.7 |
| 19/01/2002 05:00 | 37.81 | 21.08 | 562 | 5 | 3.2 | 10.1 | 28/01/2002 11:42 | 37.13 | 21.09 | 631 | 5 | 2.9 | 7.3 |
| 19/01/2002 06:56 | 39.15 | 24.17 | 376 | 28 | 3.0 | 21.2 | 28/01/2002 21:02 | 38.12 | 20.64 | 555 | 5 | 3.0 | 9.8 |
| 19/01/2002 07:24 | 38.23 | 23.59 | 465 | 10 | 4.2 | 19.5 | 28/01/2002 21:16 | 37.19 | 21.10 | 624 | 5 | 2.8 | 7.2 |
| 19/01/2002 08:37 | 38.16 | 20.61 | 553 | 5 | 3.2 | 10.5 | 29/01/2002 01:04 | 38.49 | 25.66 | 512 | 10 | 3.1 | 11.8 |
| 19/01/2002 09:35 | 38.69 | 23.90 | 419 | 5 | 3.0 | 17.1 | 29/01/2002 14:02 | 38.57 | 23.68 | 428 | 5 | 2.7 | 14.7 |
| 19/01/2002 10:05 | 39.13 | 24.24 | 380 | 19 | 3.4 | 23.5 | 29/01/2002 19:03 | 38.13 | 20.21 | 580 | 5 | 3.6 | 10.7 |
| 19/01/2002 15:16 | 37.96 | 20.78 | 562 | 5 | 3.2 | 10.1 | 30/01/2002 18:17 | 38.57 | 24.33 | 443 | 18 | 3.6 | 18.4 |
| 20/01/2002 00:10 | 40.61 | 32.91 | 1094 | 10 | 3.9 | 3.3 | 30/01/2002 20:03 | 37.66 | 21.66 | 554 | 5 | 4.4 | 14.3 |
| 20/01/2002 02:33 | 40.21 | 19.66 | 464 | 5 | 3.1 | 14.4 | 31/01/2002 00:43 | 34.58 | 34.01 | 1479 | 25 | 3.0 | 1.4 |
| 20/01/2002 02:38 | 38.49 | 24.34 | 452 | 16 | 3.3 | 16.2 | 31/01/2002 10:16 | 34.62 | 31.96 | 1299 | 25 | 3.0 | 1.8 |
| 20/01/2002 12:27 | 38.47 | 25.34 | 496 | 43 | 3.5 | 14.3 | 31/01/2002 10:30 | 37.54 | 21.84 | 561 | 5 | 3.6 | 11.5 |

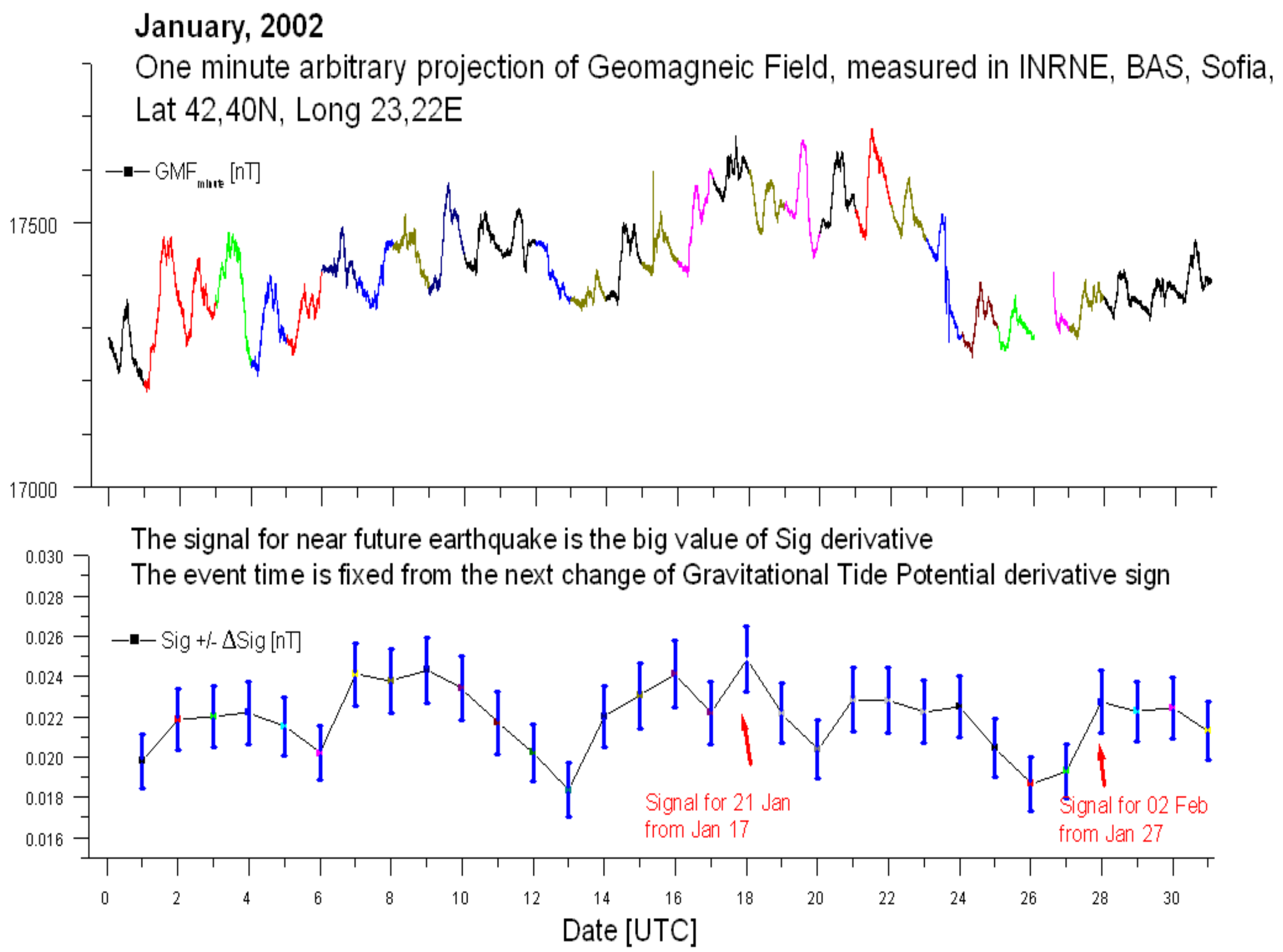

Figure 3

If the many points digital derivative dSig/dt of Sig has unusual big value, as it was on January 18 (See Figure 3, down picture one has a signal for a near future earthquake in the region. The day of the event is determined by the near future minimum or maximum of the Tide Gravitational Potential behavior (Venedikov at all [7], Figure 4.), averaged for the day. The predicted time interval is +/- 1 day in case of minimum and +/-2 days in case of maximum.

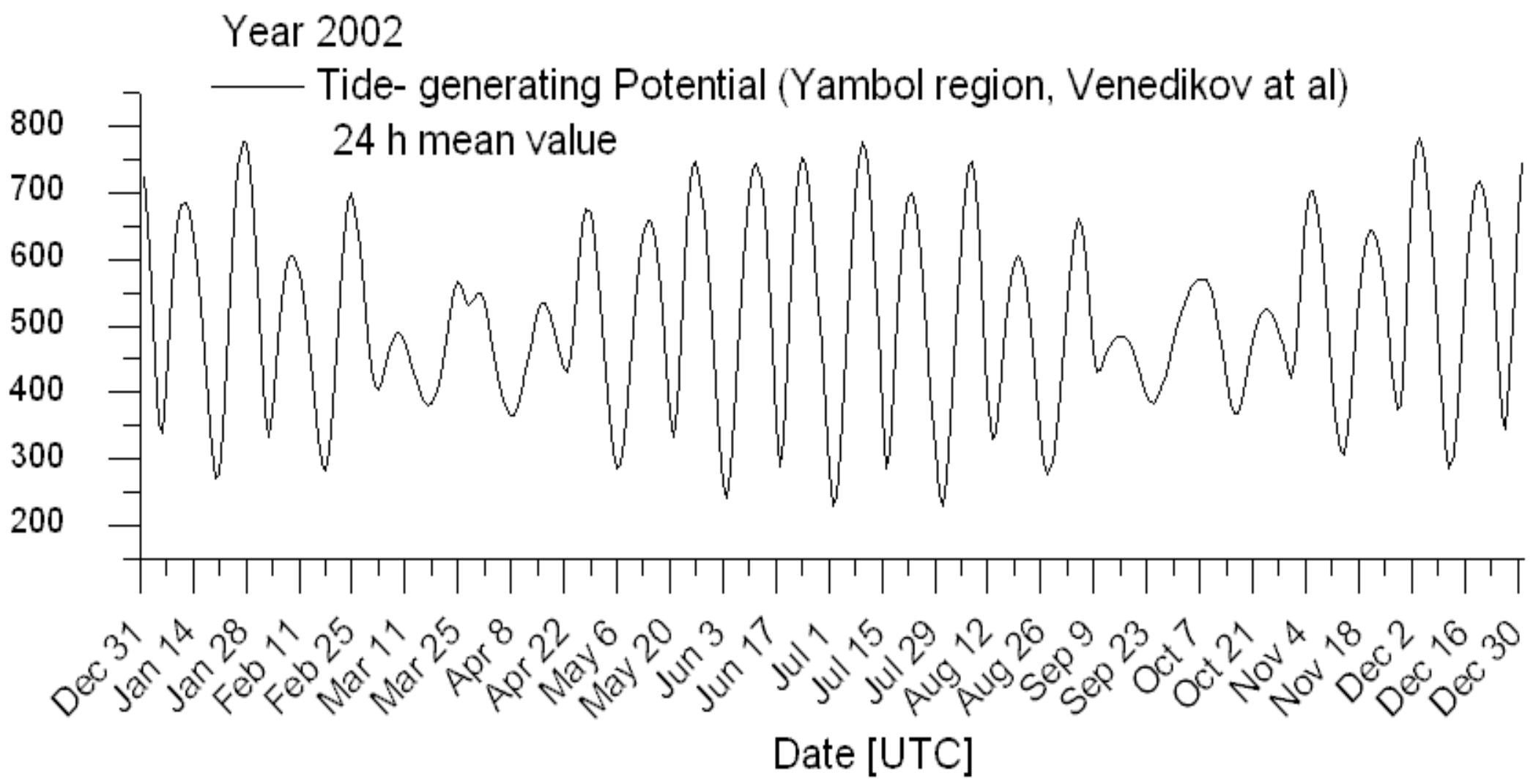

Figure 4

The days after Signal we have to take into account the possible influence of the Sun, analyzing the information for the Space weather [8] (http://www.sec.noaa.gov/SWN/), the 3-daySatelliteEnvironment from SEK [9], (http://www.sec.noaa.gov/rt_plots/satenv.html), the 3- day GOES 8&10 X-ray Flux [10], (http://www.sec.noaa.gov/rt_plots/xray_5m.html), all other possible core parameters, and of course the real time earthquake information from European- Mediterranean Seismological Centre [11] (http://www.emsc-csem.org/).

Additional confirmation that unusual big value of dSig/dt is a signal for near time earthquake in the region is the variation of $GMF_m$ with 3- 10 minutes period around the usual daily behavior and amplitude less than 10 – 20 nT.

At the end of the month, when the Earthquakes information is collected from USGS, Earthquake Hazards program, National Earthquake Information Center, World Data Center for Seismology, Denver [12], (http://wwwneic.cr.usgs.gov/neis/bulletin/), from the Earthquake Map (Figure 5) and Earthquake Table (Table 1) for the region we can test the reliability of the prediction (Figure 6).

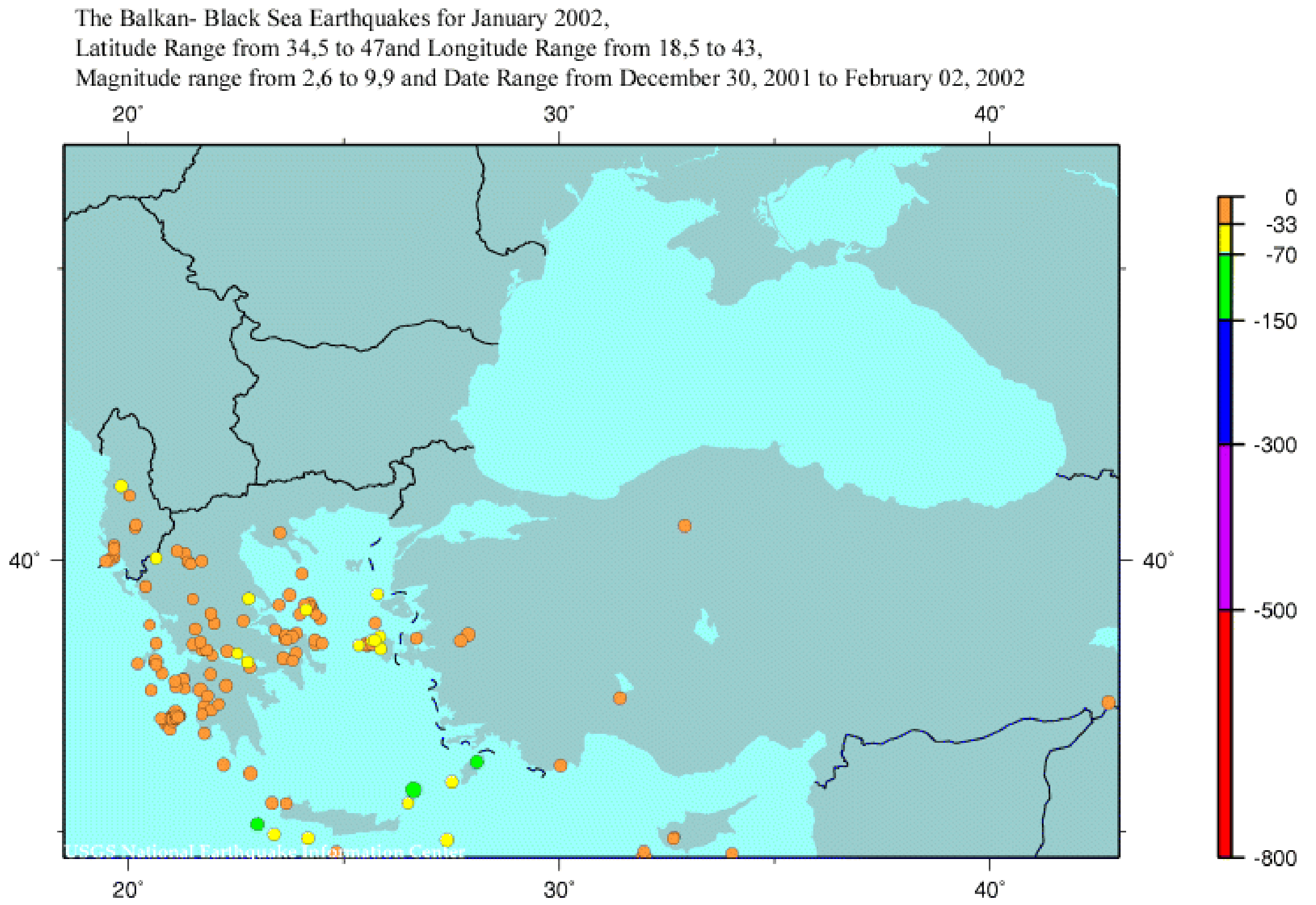

Figure 5 January 2002 Balkan- Black Sea Earthquakes Map

Table 1.The earthquake data for January, 2002. The data for predicted events are bold.

The main part of the test is focused on the behavior of the new parameter SChtM. It is proportional to the Richter magnitude of the event, inversely proportional to the squared distances in kilometers, divided by 1000:

$$SChtM = Magnitude/(Km/1000)^2 .$$

This new parameter SChtM appears as a natural measure of the transferred energy from the epicenter to the measurement point at the moment of geomagnetic quake, the anomalous behavior of which is a signal for a near future Earthquake.

As everyone can see from Table 1, the events with relatively big value of the new parameter are in the time window correlated with the time predicted from the daily Signal curve from down picture of Figure 3. The difference is in the framework of time tolerance +/- 1 day or +/- 2 days, depending on the minimum or maximum of tide potential. The bigger time deviation in some cases is connected with the fact that the time of minimums or maximums of the tide is with accuracy of 1 day. The hour tide analysis should decrease the deviation, but at this stage of research such accuracy seems enough well.

The next Figure 6 illustrates the time behavior of the new parameter SChtM:

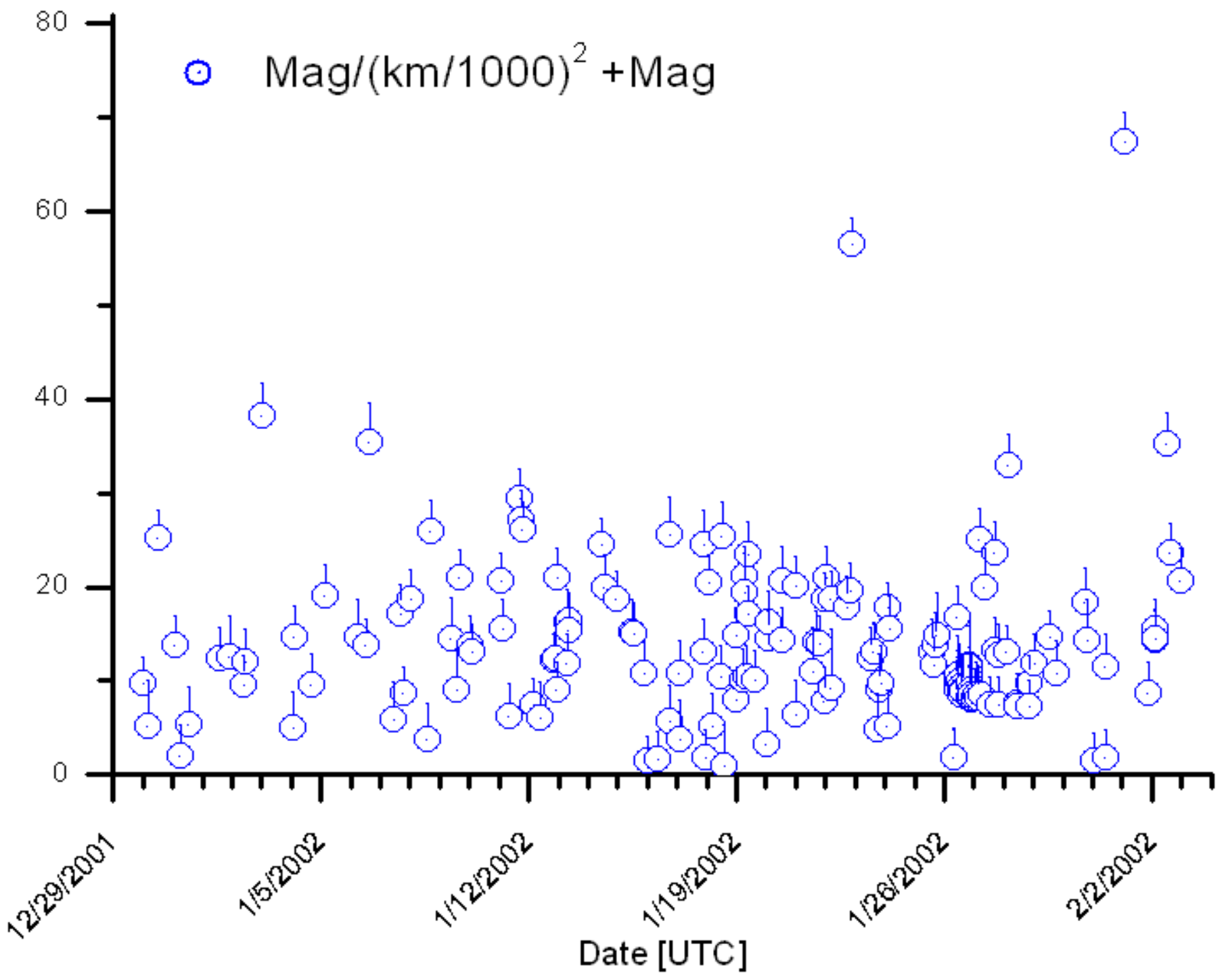

Figure 6

So, one can say, that the parameter SChtM separates the earthquakes, which incoming has been registered and predicted by the geomagnetic fields behavior, from all other earthquakes, which take place in the region at the same time period. Their Magnitude has not been "so big", the distance from the measurement point has been "big enough", and the quality of the Earth crust has been such that its influence has not been detected visibly with our device.

The step by step formal mathematical modeling and a deeper physical understanding of the integrity of these intrinsically connected processes (relative mass movement, energy and electrical distributions and currents, Sun wind influence, etc.) through making use of the inverse problem methods will bring us to a clarification of the physical meaning of "big enough" and "so far away", to corresponding numerical estimations, to discoveries of functional dependences of the local crust dynamical state parameters on other appropriate environmental parameters of Earth, Atmosphere, Near Space and Sun.

## The statistics from January to June 2002

The figures 7-9 bellow give a graphical presentation of our attempt to make a Statistical estimation for earthquake predictions for the Balkan- Black sea region based on the measurement of a certain projection of the geomagnetic field for the period January- June 2002.

In Figure 7 the time of the geomagnetic signal for near future earthquake is represented (star). The estimated time interval is represented by box, and the circles are the values of new parameter. It is seen that all earthquakes are in the boxes. Figure 7 represents the distribution of the deviation of the prediction with step of 4 hours. Figure 8 represents the deviation data.

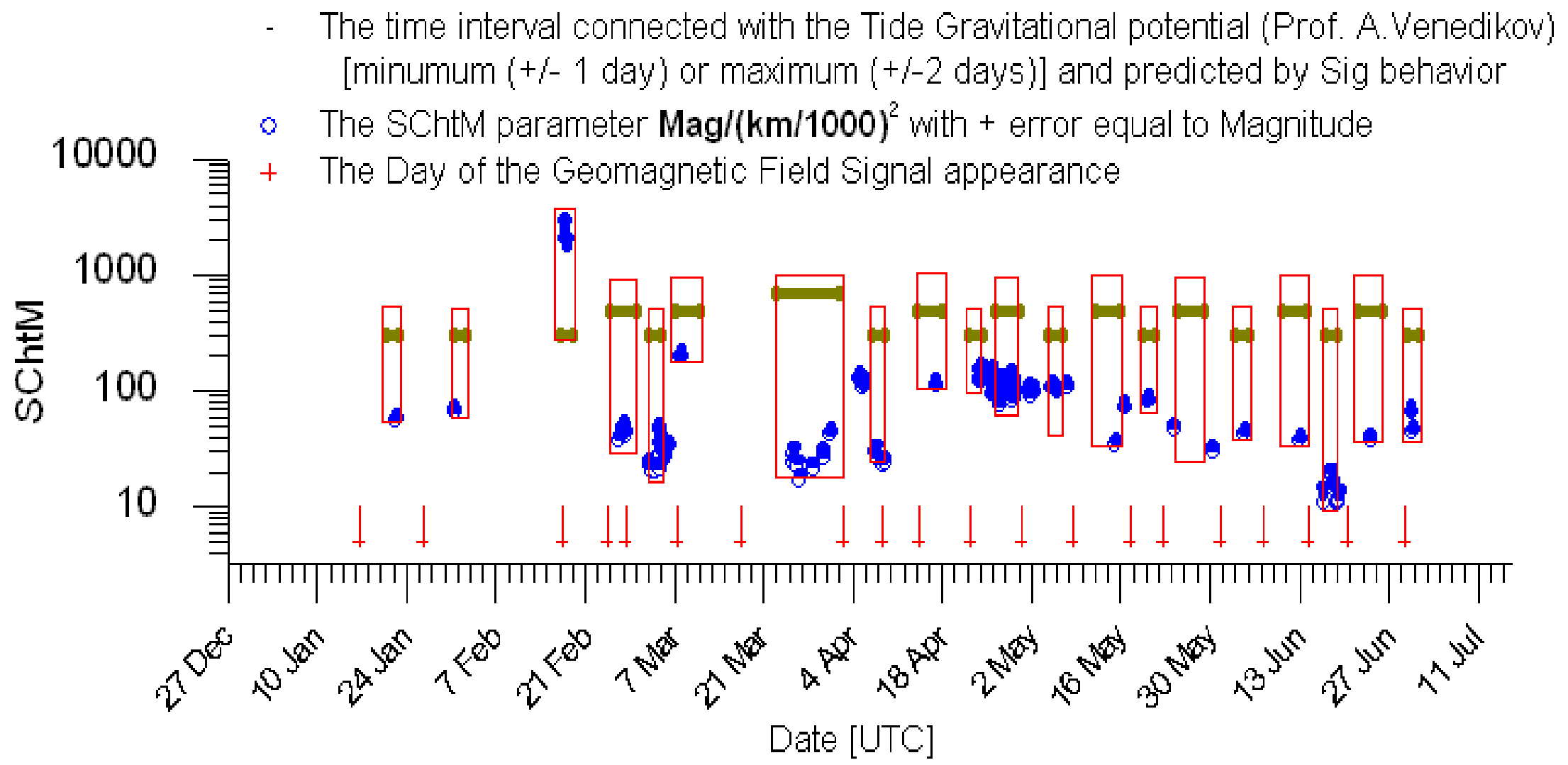

Figure 7

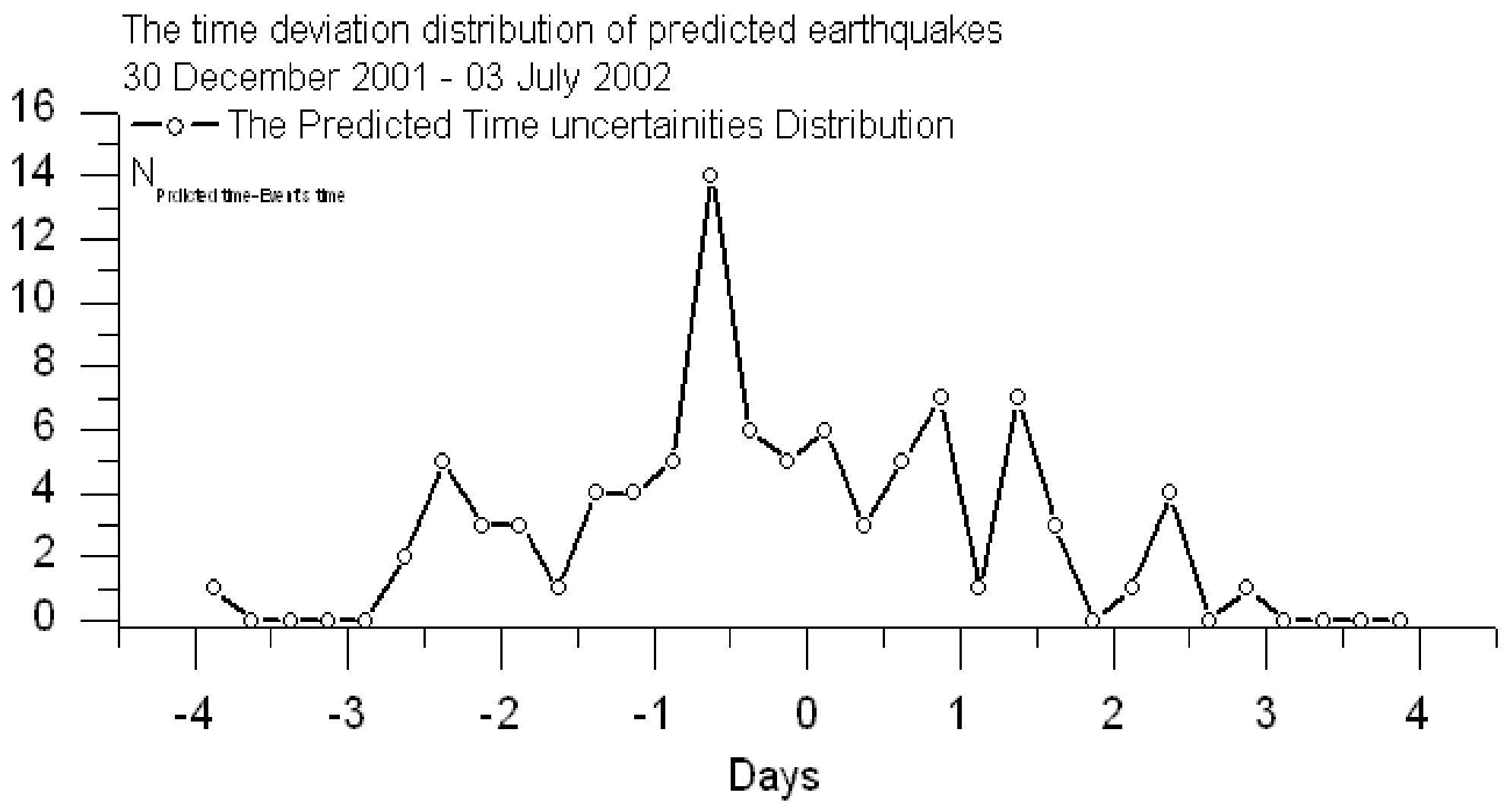

Figure 8

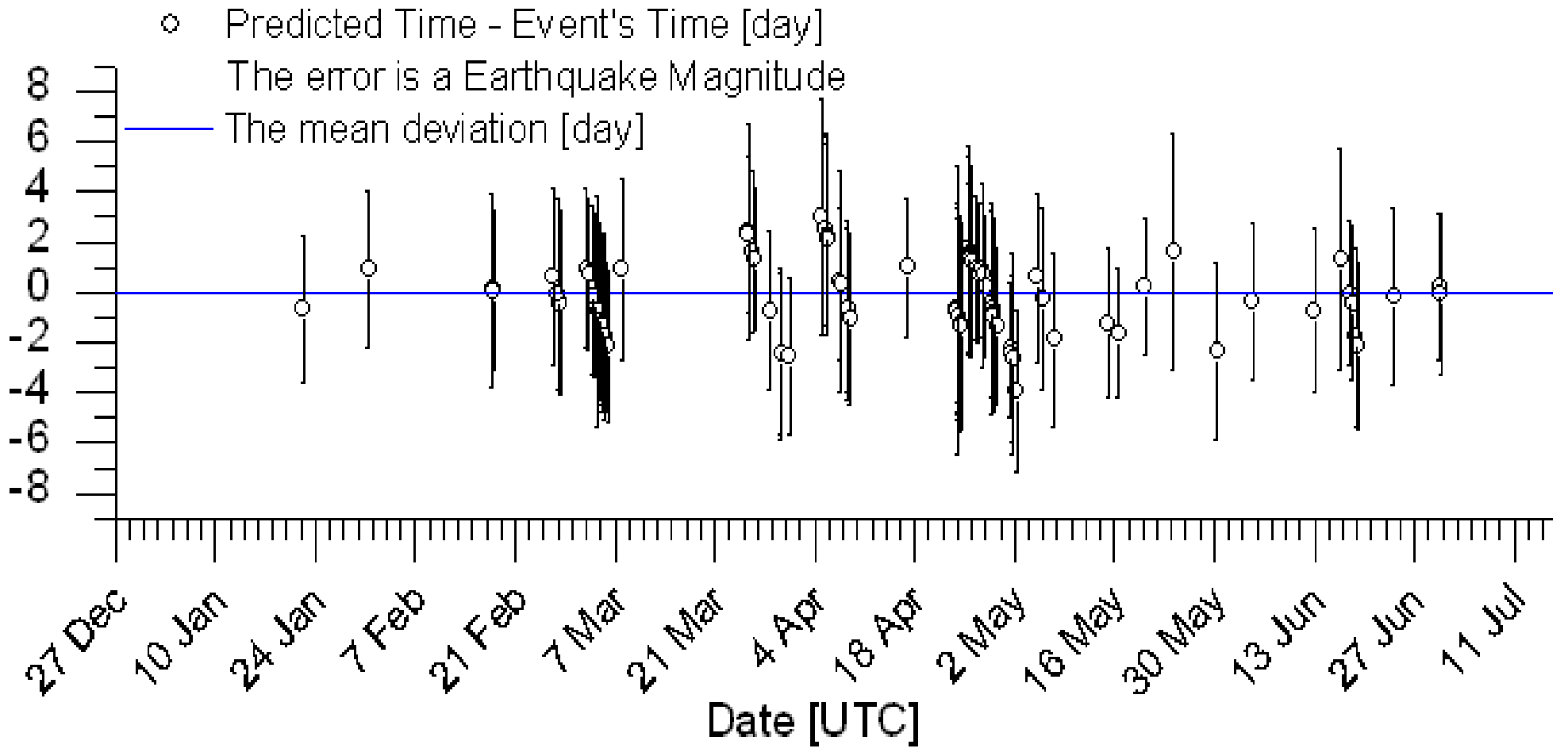

Figure 9

The problem of “large” earthquake prediction will be analyzed in next papers on the basis of Fourier analysis of the GMF daily behavior. The correlation with the Sun wind influence (“Sun noise”) will allow to connect the unusual new frequencies and the time behavior of its amplitudes with the type (point, line, etc.) and magnitude of the near future event. From such point of view the fore- and after shock earthquakes have to be the same event. This is a possible explanation why we have more earthquakes than prediction signals.

The next Table 2 gives the earthquakes forecast time characteristics for January- June, 2002 period.

Table 2

| Number | Date of Signal | Event's time from | Event's time to | Tide | Forecast days |
|---|---|---|---|---|---|
| 1 | Jan 17 | Jan 20 | Jan 22 | min | 3 |
| 2 | Jan 27 | Feb 01 | Feb 03 | min | 4 |
| 3 | Jan 31 | Feb 08 | Feb 10 | min | 8 |
| 4 | Feb 08 | Feb 16 | Feb 18 | min | 5 |
| 5 | Feb 18 | Feb 23 | Feb 27 | max | 2 |
| 6 | Feb 28 | Mar 02 | Mar 04 | min | 8 |
| 7 | Mar 08 | Mar 16 | Mar 18 | min | 5 |
| 8 | Mar 18 | Mar 23 | Apr 01 | max, min, max | 4 |
| 9 | Apr 03 | Apr 07 | Apr 09 | min | 5 |
| 10 | Apr 08 | Apr 13 | Apr 17 | max | 7 |
| 11 | Apr 15 | Apr 22 | Apr 24 | min | 3 |
| 12 | Apr 23 | Apr 26 | Apr 30 | max | 4 |
| 13 | May 01 | May 05 | May 07 | min | 3 |
| 14 | May 09 | May 12 | May 16 | max | 1 |
| 15 | May 18 | May 19 | May 21 | min | 1 |
| 16 | May 23 | May 24 | May 28 | max | 1 |
| 17 | Jun 01 | Jun 03 | Jun 05 | min | 2 |
| 18 | Jun 08 | Jun 09 | Jun 13 | max | 1 |
| 19 | Jun 15 | Jun 17 | Jun 19 | min | 2 |
| 20 | Jun 21 | Jun 22 | Jun 26 | max | 1 |
| 21 | Jun 30 | Jul 01 | Jul 03 | min | 1 |

Table 2

In the appendix are the February- July, 2002 month reports, where are included the data for earthquakes [11], from which everyone can test the correctness of the above statistics.

## Balkan- Black Sea Earthquakes Prediction NETWORK

Each stationary monitoring site has to include [2]:

- The vector (tree components) Geomagnetic field measurement (GMF3) with absolute accuracy less than 1nT, 50-100 Hz samples per second;
- The two components measurement (EE2) of earth electric current (Thanassoulas current site in Volos, Greece) with 5 –10 % relative accuracy, 50-100 Hz samples per second;
- One-dimensional vertical atmospheric electro- potential and current measurements (AE1) with 5 –10 % relative accuracy, 50-100 Hz samples per second;
- The temperature T of the Earth in depth, where is not a day and season variations with 0.1 C Degrees absolute accuracy and 1 hour sample.

The NETWORK sites must be appropriately arranged in accordance with the Earthquake Risk Geological map.

When the future epicenter area is determined by EE2, the mobile site has to start there GMF3 and AE1 measurements. All other possible geological, geo-chemical and geo-physical and biological precursor parameters have to be included in the analysis. The Magnitude of the expected earthquake will depend on time and space gradients of the parameters. This unknown function will be found after some time for obtaining of “enough” statistics and using the inverse problem methods.

## Conclusion

The discovery of the parameter SChtM, which is some measure of the earthquake “energy” influence at the measurement points, allows accepting the geomagnetic quake as a reliable precursor of near future earthquake event. The other earthquakes in the same period have to be very “slow” or very “far away”, or very “slow” and “far away”.

The relative maximums of this parameter is clearly correlated with the extreme time points of the tide gravitational potential, which can be seen from the distribution of the deviation between predicted and occurred earthquake time.

The proposed NETWORK will help to clarify this correlation, because its first result will be the “when, where” prediction of the earthquake time and epicenter. The time behavior of vertical electro-potential and Earth temperature will give “how strong” prediction for the future earthquake. Obviously, taking into account the geological, geo-chemical, geo-physical and biological earthquake precursors will be essential for the prediction accuracy.

Summarizing, one can say:

- The geomagnetic quake is a reliable precursor for a near future earthquake;

- In the Balkan- Black Sea region the earthquake time is fixed by the extreme points of the Tide gravitational potential
- The problem "where" and "how strong" can be treated after, at least, one site of the proposed Balkan- Black Sea Network will start to operate.

This approach can be tested with the historical Geomagnetic and Earthquake data.

Acknowledgements

The author would like to note that the collaboration with C. Thanassoulas in the time of Sofia, July, 2002 Seminar allowed to make the first Earthquake "when and where" prediction- 26 July, 2002 Skyros, Greece earthquake.

It is a pleasure to mention the technical software and hardware assistance of I. Angelov, I. Kalapov, Cht.S. Mavrodiev and R. Ugrinov, the interest and fruitful discussions with E.B. Alexandrov, E. Botev, A. Donkov, S. Donev, V.G. Kadyshevsky, B. Rangelov, A.N. Sissakian, A.N. Skrinsky, A. Venedikov, the members of LTPh, JINR, Dubna, INRNE, BAS, GPHI, BAS seminars on this topic and the support of the INRNE Director J. Stamenov.

**APPENDIX**

# February 2002 Report

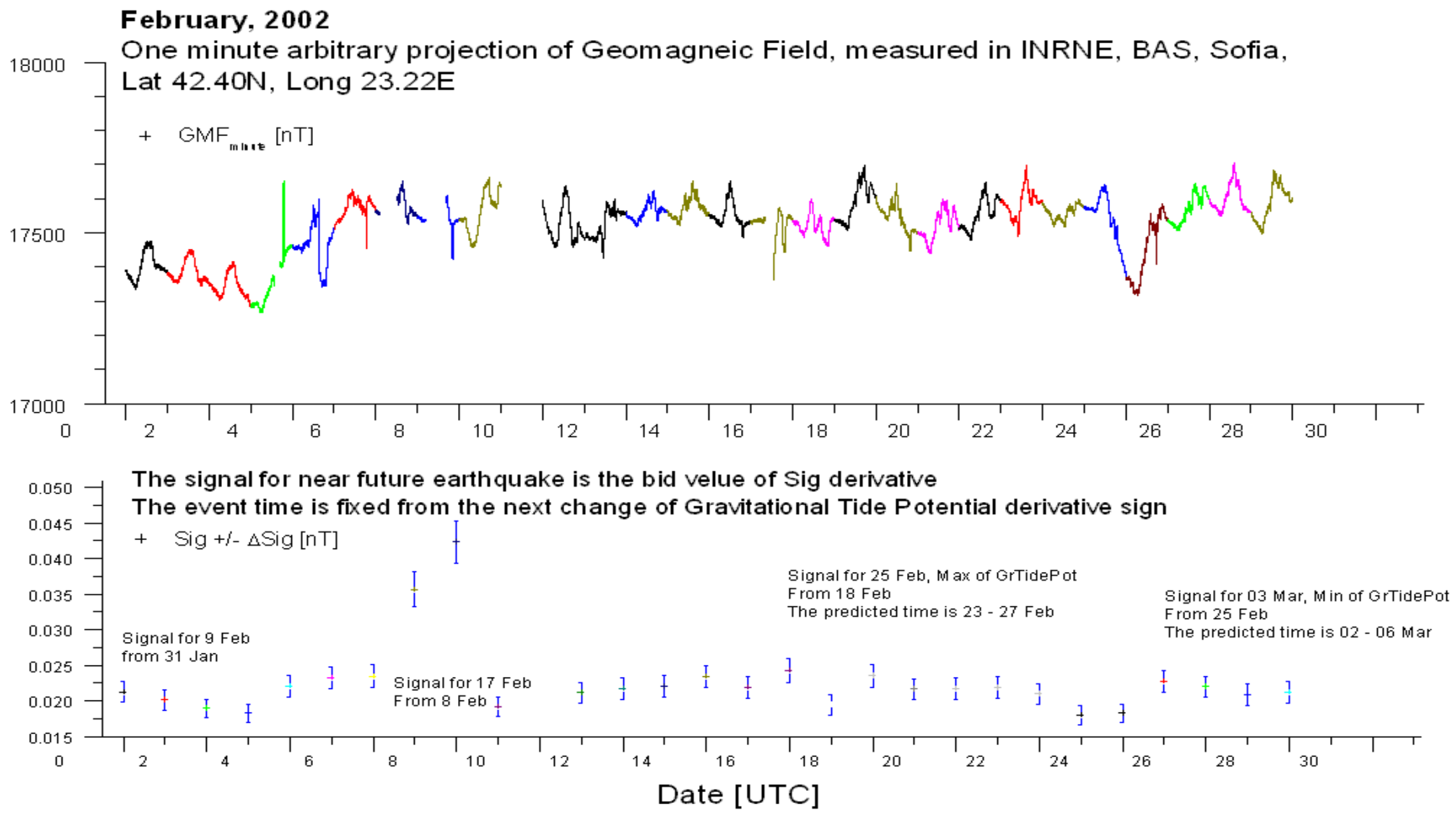

Figure February2002GMFSignal

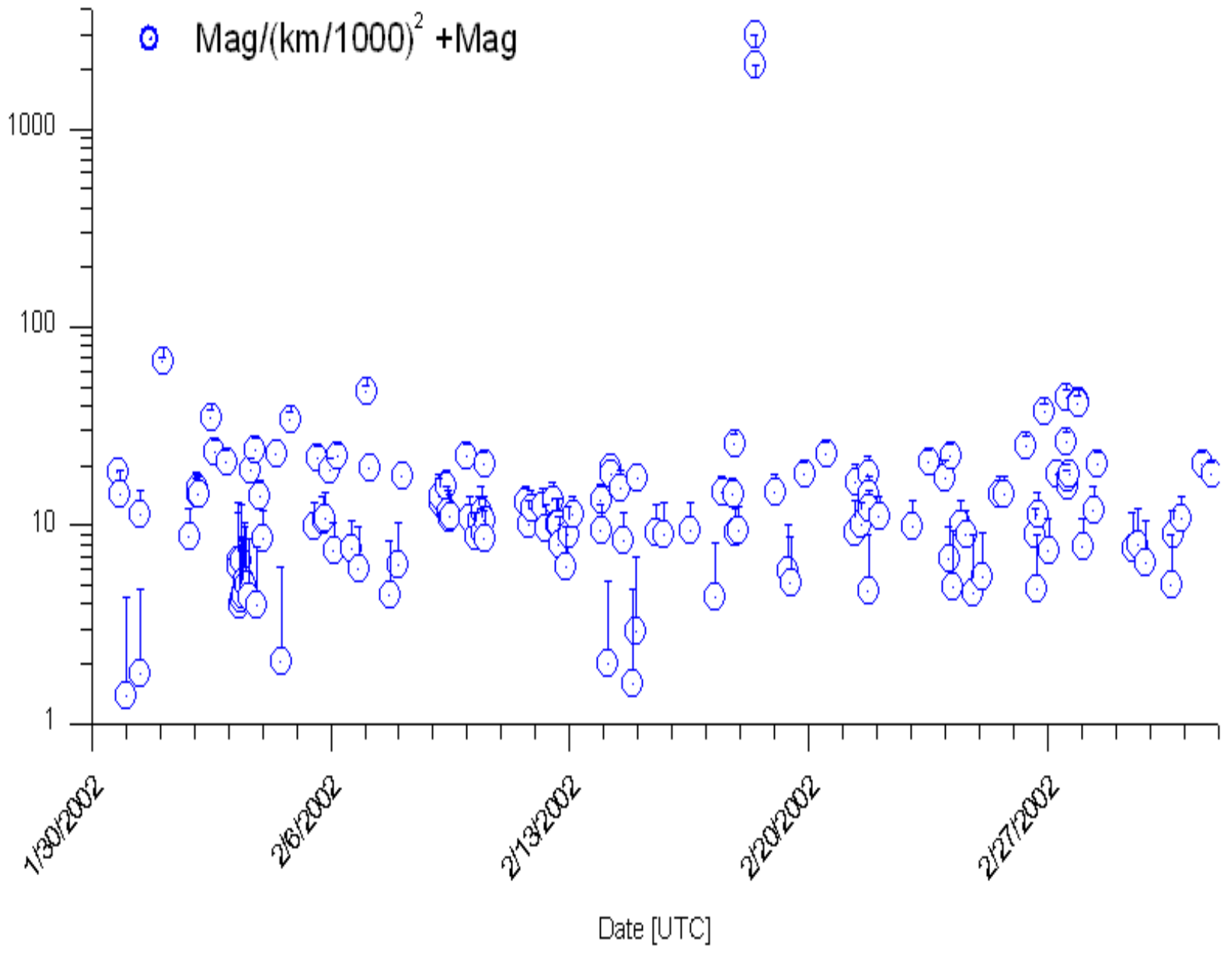

Figure February2002,SChtM

| EQRable2002 | Feb | | | | | | | | | | | | |
|---|---|---|---|---|---|---|---|---|---|---|---|---|---|
| ddmmyy hhmm | Lat | Lon | kmSof | Dep | Mag | SChtM | ddmmyy hhmm | Lat | Lon | kmSof | Dep | Mag | SChtM |
| 01/02/2002 02:12 | 40.49 | 23.51 | 214 | 8 | 3.1 | 67.4 | 12/02/2002 23:13 | 37.12 | 20.98 | 637 | 5 | 3.6 | 8.9 |
| 01/02/2002 20:47 | 37.15 | 21.15 | 626 | 5 | 3.4 | 8.7 | 13/02/2002 02:31 | 38.08 | 22.86 | 481 | 5 | 2.6 | 11.2 |
| 02/02/2002 02:24 | 38.49 | 21.49 | 475 | 5 | 3.2 | 14.2 | 13/02/2002 22:05 | 37.63 | 20.78 | 595 | 5 | 3.3 | 9.3 |
| 02/02/2002 02:28 | 38.50 | 24.49 | 455 | 5 | 3.2 | 15.4 | 13/02/2002 22:11 | 38.41 | 21.92 | 466 | 5 | 2.9 | 13.4 |
| 02/02/2002 03:21 | 38.53 | 21.66 | 463 | 27 | 3.1 | 14.5 | 14/02/2002 03:04 | 35.74 | 32.46 | 1264 | 25 | 3.2 | 2.0 |
| 02/02/2002 12:10 | 39.76 | 24.02 | 306 | 5 | 3.3 | 35.2 | 14/02/2002 04:44 | 38.60 | 23.62 | 424 | 5 | 3.5 | 19.5 |
| 02/02/2002 15:02 | 39.20 | 24.08 | 368 | 14 | 3.2 | 23.7 | 14/02/2002 05:37 | 38.64 | 23.96 | 425 | 5 | 3.3 | 18.2 |
| 02/02/2002 23:05 | 38.75 | 23.41 | 406 | 10 | 3.4 | 20.7 | 14/02/2002 11:46 | 38.55 | 25.53 | 498 | 10 | 3.8 | 15.3 |
| 03/02/2002 07:11 | 38.57 | 31.27 | 990 | 5 | 6.5 | 6.6 | 14/02/2002 14:22 | 36.82 | 22.94 | 620 | 54 | 3.2 | 8.3 |
| 03/02/2002 07:14 | 38.70 | 30.87 | 943 | 10 | 5.5 | 6.2 | 14/02/2002 20:31 | 35.22 | 33.72 | 1412 | 25 | 3.2 | 1.6 |
| 03/02/2002 07:57 | 38.61 | 31.21 | 982 | 10 | 3.8 | 3.9 | 14/02/2002 22:41 | 35.66 | 31.58 | 1192 | 58 | 4.1 | 2.9 |
| 03/02/2002 09:26 | 38.63 | 30.90 | 950 | 10 | 6.0 | 6.7 | 15/02/2002 00:08 | 39.42 | 20.93 | 417 | 5 | 3.0 | 17.2 |
| 03/02/2002 09:38 | 38.59 | 30.84 | 946 | 10 | 3.8 | 4.3 | 15/02/2002 13:15 | 37.16 | 21.14 | 626 | 5 | 3.6 | 9.2 |
| 03/02/2002 09:55 | 38.62 | 30.76 | 936 | 10 | 4.0 | 4.6 | 15/02/2002 18:38 | 36.44 | 22.23 | 671 | 62 | 4.0 | 8.9 |
| 03/02/2002 10:00 | 38.58 | 31.22 | 984 | 10 | 4.3 | 4.4 | 16/02/2002 13:12 | 37.60 | 20.60 | 607 | 5 | 3.5 | 9.5 |
| 03/02/2002 10:15 | 38.51 | 30.87 | 953 | 10 | 4.0 | 4.4 | 17/02/2002 06:23 | 38.55 | 30.88 | 952 | 33 | 3.9 | 4.3 |
| 03/02/2002 11:39 | 38.55 | 31.18 | 982 | 10 | 5.1 | 5.3 | 17/02/2002 11:42 | 39.61 | 26.40 | 470 | 28 | 3.2 | 14.5 |
| 03/02/2002 11:54 | 38.59 | 31.17 | 979 | 10 | 4.8 | 5.0 | 17/02/2002 19:17 | 40.24 | 19.61 | 467 | 5 | 3.1 | 14.2 |
| 03/02/2002 14:40 | 38.59 | 31.32 | 994 | 10 | 4.3 | 4.4 | 17/02/2002 20:12 | 41.02 | 20.30 | 359 | 5 | 3.3 | 25.7 |
| 03/02/2002 15:47 | 38.71 | 23.71 | 413 | 21 | 3.2 | 18.7 | 17/02/2002 20:33 | 37.65 | 21.47 | 562 | 5 | 2.9 | 9.2 |
| 03/02/2002 19:06 | 39.14 | 24.14 | 376 | 30 | 3.4 | 24.1 | 17/02/2002 23:15 | 37.66 | 21.42 | 563 | 5 | 3.0 | 9.5 |
| 03/02/2002 20:31 | 38.48 | 31.08 | 975 | 10 | 3.8 | 4.0 | 18/02/2002 10:19 | 42.08 | 23.15 | 36 | 33 | 3.9 | 2950.0 |
| 03/02/2002 22:31 | 40.93 | 19.23 | 472 | 8 | 3.1 | 13.9 | 18/02/2002 10:44 | 42.06 | 23.12 | 39 | 33 | 3.2 | 2067.8 |
| 04/02/2002 00:49 | 37.19 | 21.17 | 622 | 5 | 3.3 | 8.5 | 19/02/2002 00:44 | 38.14 | 23.66 | 475 | 8 | 3.3 | 14.6 |
| 04/02/2002 10:02 | 38.80 | 23.00 | 400 | 33 | 3.7 | 23.1 | 19/02/2002 10:12 | 36.14 | 27.78 | 860 | 33 | 4.3 | 5.8 |
| 04/02/2002 13:06 | 40.01 | 35.70 | 1411 | 10 | 4.1 | 2.1 | 19/02/2002 11:35 | 35.66 | 26.96 | 856 | 5 | 3.7 | 5.1 |
| 04/02/2002 19:45 | 41.30 | 20.59 | 316 | 5 | 3.4 | 34.0 | 19/02/2002 21:17 | 38.95 | 24.30 | 401 | 15 | 2.9 | 18.0 |
| 05/02/2002 12:37 | 37.65 | 20.84 | 590 | 5 | 3.4 | 9.8 | 20/02/2002 12:46 | 38.73 | 23.46 | 408 | 7 | 3.8 | 22.8 |
| 05/02/2002 14:46 | 39.17 | 24.06 | 371 | 39 | 3.0 | 21.9 | 21/02/2002 08:40 | 36.62 | 21.59 | 667 | 41 | 4.1 | 9.2 |
| 05/02/2002 18:53 | 38.09 | 20.64 | 558 | 5 | 3.3 | 10.6 | 21/02/2002 09:15 | 38.39 | 21.77 | 473 | 5 | 3.7 | 16.5 |
| 05/02/2002 20:18 | 37.75 | 20.78 | 583 | 5 | 3.7 | 10.9 | 21/02/2002 13:25 | 37.68 | 21.80 | 547 | 5 | 3.0 | 10.0 |
| 05/02/2002 23:42 | 38.87 | 22.57 | 398 | 37 | 3.0 | 18.9 | 21/02/2002 18:04 | 38.67 | 30.91 | 949 | 33 | 4.2 | 4.7 |
| 06/02/2002 02:11 | 37.11 | 21.03 | 636 | 5 | 3.0 | 7.4 | 21/02/2002 18:20 | 38.37 | 21.79 | 475 | 10 | 4.1 | 18.2 |
| 06/02/2002 05:07 | 39.72 | 25.33 | 379 | 18 | 3.2 | 22.3 | 21/02/2002 18:29 | 38.34 | 21.80 | 477 | 5 | 3.3 | 14.5 |
| 06/02/2002 15:02 | 37.17 | 21.08 | 627 | 5 | 3.0 | 7.6 | 21/02/2002 18:47 | 38.36 | 21.70 | 479 | 9 | 2.8 | 12.2 |
| 06/02/2002 20:01 | 35.37 | 23.43 | 781 | 5 | 3.7 | 6.1 | 22/02/2002 02:01 | 37.98 | 22.32 | 501 | 25 | 2.8 | 11.2 |
| 07/02/2002 01:22 | 40.31 | 23.99 | 247 | 11 | 2.9 | 47.4 | 23/02/2002 01:06 | 37.87 | 26.42 | 616 | 5 | 3.7 | 9.8 |
| 07/02/2002 03:39 | 38.60 | 23.71 | 425 | 6 | 3.5 | 19.4 | 23/02/2002 12:35 | 43.25 | 19.92 | 378 | 10 | 3.0 | 21.0 |
| 07/02/2002 17:53 | 36.20 | 28.95 | 937 | 35 | 3.9 | 4.4 | 24/02/2002 00:31 | 38.32 | 21.80 | 480 | 10 | 4.0 | 17.4 |
| 07/02/2002 23:34 | 35.41 | 24.23 | 784 | 8 | 3.9 | 6.4 | 24/02/2002 02:48 | 36.65 | 21.61 | 663 | 10 | 3.0 | 6.8 |
| 08/02/2002 02:19 | 38.99 | 25.10 | 432 | 26 | 3.3 | 17.7 | 24/02/2002 04:16 | 39.10 | 24.21 | 382 | 18 | 3.3 | 22.6 |
| 09/02/2002 04:43 | 38.03 | 20.83 | 553 | 5 | 4.0 | 13.1 | 24/02/2002 05:32 | 34.81 | 27.28 | 955 | 68 | 4.5 | 4.9 |
| 09/02/2002 04:55 | 38.08 | 20.84 | 548 | 5 | 4.2 | 14.0 | 24/02/2002 11:14 | 37.88 | 20.92 | 563 | 5 | 3.2 | 10.1 |
| 09/02/2002 09:51 | 38.64 | 21.48 | 460 | 5 | 3.3 | 15.6 | 24/02/2002 14:50 | 37.59 | 21.38 | 572 | 22 | 2.9 | 8.9 |
| 09/02/2002 10:25 | 37.60 | 20.86 | 594 | 5 | 4.1 | 11.6 | 24/02/2002 19:41 | 38.45 | 31.13 | 981 | 10 | 4.4 | 4.6 |
| 09/02/2002 11:00 | 37.53 | 20.73 | 607 | 5 | 4.0 | 10.9 | 25/02/2002 02:23 | 35.64 | 25.90 | 807 | 7 | 3.6 | 5.5 |
| 09/02/2002 12:29 | 38.48 | 20.50 | 530 | 5 | 3.1 | 11.1 | 25/02/2002 14:41 | 39.35 | 26.28 | 480 | 5 | 3.3 | 14.4 |
| 09/02/2002 23:40 | 39.13 | 22.46 | 373 | 18 | 3.1 | 22.3 | 25/02/2002 17:45 | 39.35 | 26.29 | 480 | 5 | 3.3 | 14.3 |
| 10/02/2002 02:12 | 37.78 | 21.92 | 533 | 17 | 3.1 | 10.9 | 26/02/2002 08:57 | 40.94 | 20.34 | 358 | 10 | 3.2 | 24.9 |
| 10/02/2002 06:12 | 37.37 | 21.58 | 587 | 5 | 3.0 | 8.7 | 26/02/2002 15:28 | 38.10 | 20.30 | 577 | 5 | 3.0 | 9.0 |
| 10/02/2002 08:42 | 37.71 | 20.96 | 578 | 5 | 3.5 | 10.5 | 26/02/2002 16:30 | 38.69 | 30.98 | 955 | 10 | 4.3 | 4.7 |
| 10/02/2002 10:09 | 37.78 | 20.99 | 569 | 5 | 3.0 | 9.3 | 26/02/2002 17:13 | 38.37 | 20.67 | 529 | 5 | 3.2 | 11.4 |
| 10/02/2002 10:11 | 37.62 | 20.95 | 587 | 10 | 4.0 | 11.6 | 26/02/2002 21:53 | 41.10 | 20.80 | 305 | 10 | 3.5 | 37.6 |
| 10/02/2002 12:09 | 38.85 | 23.75 | 398 | 5 | 3.2 | 20.2 | 26/02/2002 22:08 | 41.09 | 20.80 | 306 | 10 | 3.5 | 37.5 |
| 10/02/2002 12:34 | 37.66 | 20.89 | 586 | 5 | 3.6 | 10.5 | 27/02/2002 00:45 | 36.34 | 22.47 | 678 | 37 | 3.4 | 7.4 |
| 10/02/2002 12:38 | 37.58 | 20.56 | 611 | 5 | 3.2 | 8.6 | 27/02/2002 06:18 | 38.64 | 23.57 | 419 | 2 | 3.2 | 18.2 |
| 11/02/2002 17:17 | 38.36 | 21.88 | 473 | 5 | 2.9 | 13.0 | 27/02/2002 13:38 | 41.09 | 20.54 | 331 | 7 | 2.9 | 26.5 |
| 11/02/2002 19:38 | 37.63 | 20.84 | 592 | 5 | 3.5 | 10.0 | 27/02/2002 13:39 | 41.17 | 20.89 | 293 | 5 | 3.8 | 44.4 |
| 11/02/2002 21:01 | 38.26 | 21.48 | 499 | 5 | 3.0 | 12.1 | 27/02/2002 13:59 | 38.60 | 24.62 | 450 | 36 | 3.2 | 15.8 |
| 12/02/2002 05:31 | 38.26 | 21.46 | 499 | 5 | 3.1 | 12.4 | 27/02/2002 14:13 | 38.67 | 24.70 | 445 | 35 | 3.4 | 17.1 |
| 12/02/2002 07:31 | 38.25 | 20.38 | 558 | 5 | 3.0 | 9.6 | 27/02/2002 15:52 | 38.68 | 24.63 | 442 | 33 | 3.5 | 18.0 |
| 12/02/2002 12:36 | 38.31 | 21.50 | 493 | 5 | 3.2 | 13.2 | 27/02/2002 21:40 | 41.00 | 20.98 | 293 | 5 | 3.6 | 41.9 |
| 12/02/2002 14:45 | 37.67 | 20.63 | 599 | 5 | 3.6 | 10.1 | 27/02/2002 21:47 | 41.11 | 20.90 | 295 | 5 | 3.6 | 41.5 |
| 12/02/2002 16:26 | 37.61 | 20.76 | 598 | 5 | 3.6 | 10.1 | 28/02/2002 01:00 | 37.33 | 20.67 | 630 | 5 | 3.1 | 7.8 |
| 12/02/2002 16:54 | 37.55 | 20.55 | 615 | 5 | 3.0 | 7.9 | 28/02/2002 08:37 | 40.82 | 28.12 | 572 | 9 | 3.9 | 11.9 |
| 12/02/2002 21:43 | 35.60 | 24.38 | 766 | 15 | 3.6 | 6.1 | 28/02/2002 11:26 | 38.79 | 22.68 | 405 | 5 | 3.3 | 20.1 |

# March 2002 Report

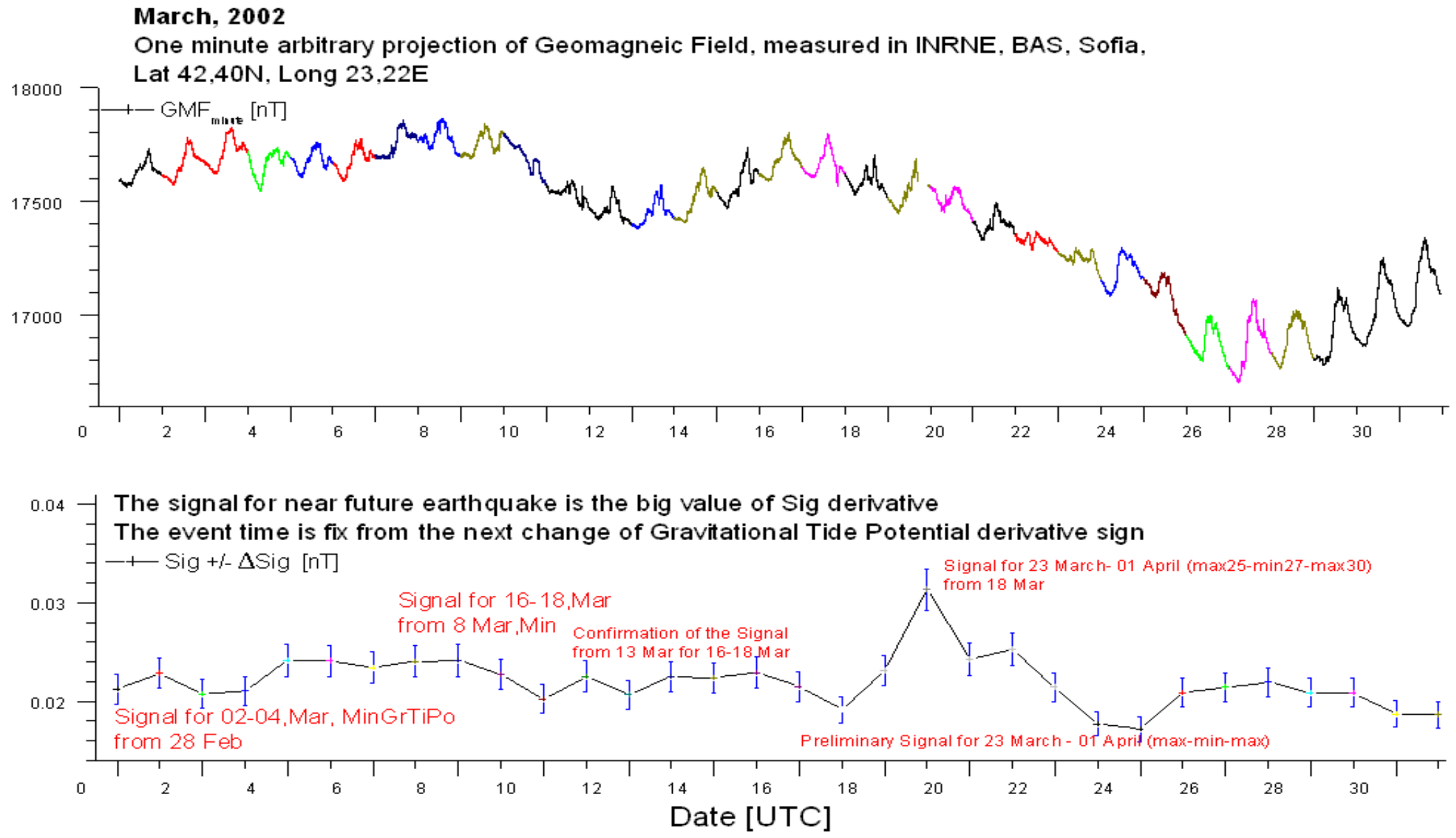

Figure March2002GMFSignal

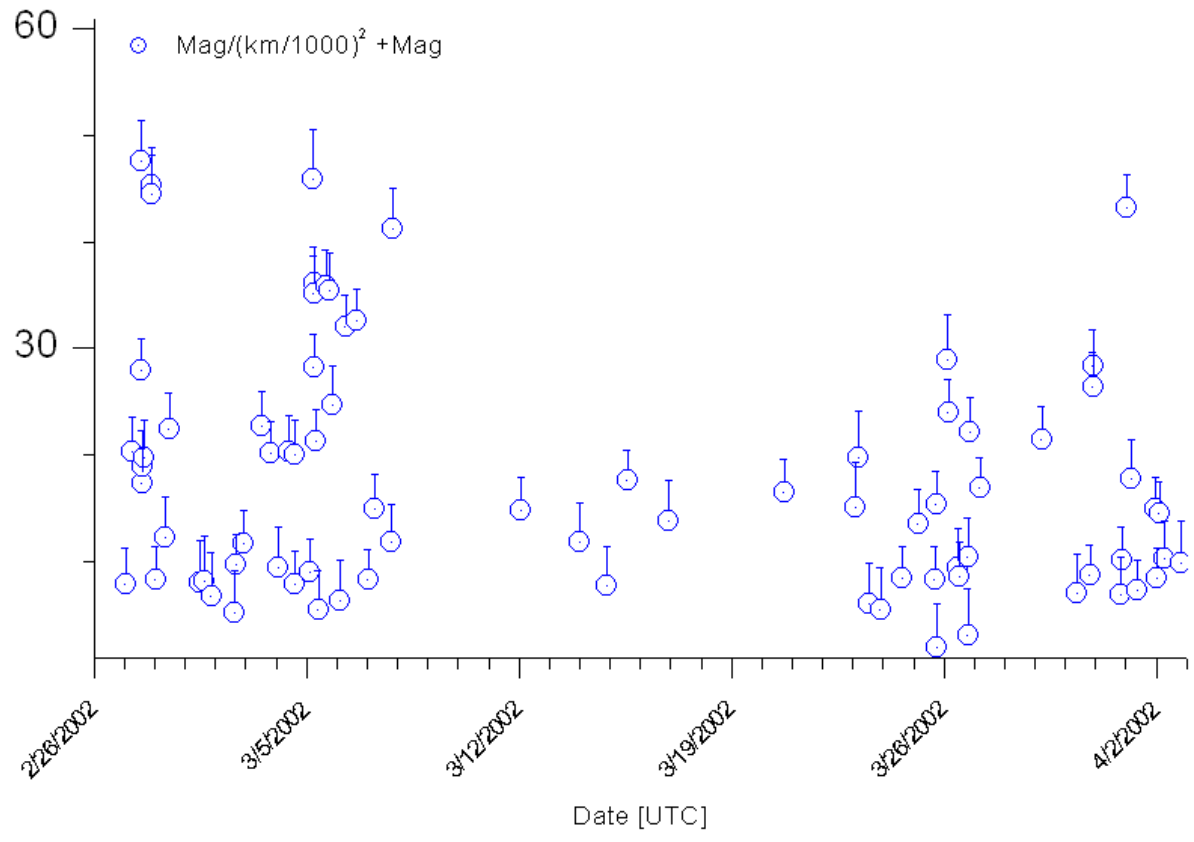

Figure March2002SChtM

**EQTable2002 March**

| ddmmyy hhmm | Lat | Lon | kmSof | Dep | Mag | SChtM | ddmmyy hhmm | Lat | Lon | kmSof | Dep | Mag | SChtM |
|---|---|---|---|---|---|---|---|---|---|---|---|---|---|
| 01/03/2002 12:01 | 38.35 | 28.31 | 708 | 10 | 4.0 | 8.0 | 12/03/2002 01:41 | 42.54 | 19.04 | 466 | 18 | 3.2 | 14.8 |
| 01/03/2002 15:47 | 38.30 | 28.24 | 706 | 10 | 4.1 | 8.2 | 14/03/2002 00:26 | 37.26 | 22.81 | 550 | 10 | 3.6 | 11.9 |
| 01/03/2002 21:09 | 37.07 | 28.00 | 778 | 33 | 4.1 | 6.8 | 14/03/2002 22:00 | 36.06 | 22.24 | 690 | 33 | 3.7 | 7.8 |
| 02/03/2002 15:10 | 35.31 | 26.81 | 862 | 25 | 3.9 | 5.2 | 15/03/2002 14:04 | 41.47 | 19.7 | 399 | 10 | 2.8 | 17.6 |
| 02/03/2002 16:20 | 37.82 | 20.96 | 547 | 5 | 2.9 | 9.7 | 16/03/2002 22:39 | 45.55 | 26.56 | 525 | 150 | 3.8 | 13.8 |
| 02/03/2002 22:32 | 38.35 | 20.63 | 515 | 5 | 3.1 | 11.7 | 20/03/2002 17:39 | 40.48 | 19.71 | 434 | 10 | 3.1 | 16.5 |
| 03/03/2002 12:50 | 38.83 | 23.46 | 375 | 22 | 3.2 | 22.8 | 23/03/2002 02:36 | 40.86 | 27.85 | 535 | 10 | 4.3 | 15.0 |
| 03/03/2002 19:45 | 38.74 | 23.56 | 386 | 5 | 3.0 | 20.1 | 23/03/2002 04:52 | 38.05 | 22.3 | 472 | 92 | 4.4 | 19.8 |
| 04/03/2002 02:06 | 37.17 | 20.71 | 624 | 5 | 3.7 | 9.5 | 23/03/2002 13:08 | 35.12 | 24.01 | 791 | 10 | 3.8 | 6.1 |
| 04/03/2002 10:16 | 38.62 | 24.11 | 410 | 21 | 3.4 | 20.3 | 23/03/2002 21:58 | 34.52 | 24.06 | 858 | 10 | 4.0 | 5.4 |
| 04/03/2002 15:15 | 38.59 | 23.81 | 406 | 25 | 3.3 | 20.0 | 24/03/2002 15:35 | 37.16 | 21.4 | 595 | 5 | 3.0 | 8.5 |
| 04/03/2002 15:18 | 37.15 | 20.73 | 625 | 5 | 3.1 | 7.9 | 25/03/2002 03:50 | 38.5 | 25.71 | 495 | 21 | 3.3 | 13.5 |
| 05/03/2002 03:06 | 36.82 | 23.07 | 597 | 6 | 3.2 | 9.0 | 25/03/2002 17:04 | 37.24 | 20.79 | 613 | 5 | 3.1 | 8.3 |
| 05/03/2002 05:23 | 40.66 | 25.62 | 317 | 19 | 4.6 | 45.9 | 25/03/2002 17:59 | 39.41 | 20.39 | 441 | 30 | 3.0 | 15.4 |
| 05/03/2002 05:52 | 40.88 | 20.84 | 302 | 5 | 3.3 | 36.2 | 25/03/2002 18:33 | 34.7 | 33.46 | 1409 | 33 | 4.0 | 2.0 |
| 05/03/2002 05:55 | 40.85 | 20.72 | 315 | 2 | 3.5 | 35.2 | 26/03/2002 03:01 | 38.93 | 24.4 | 386 | 28 | 4.3 | 28.9 |
| 05/03/2002 06:04 | 40.63 | 20.74 | 326 | 10 | 3.0 | 28.3 | 26/03/2002 03:47 | 39.1 | 24.15 | 359 | 28 | 3.1 | 24.0 |
| 05/03/2002 07:45 | 38.88 | 23.88 | 376 | 51 | 3.0 | 21.3 | 26/03/2002 11:37 | 37.01 | 21.03 | 625 | 10 | 3.7 | 9.5 |
| 05/03/2002 09:18 | 35.59 | 26.56 | 822 | 20 | 3.7 | 5.5 | 26/03/2002 12:41 | 37.14 | 20.87 | 619 | 5 | 3.3 | 8.6 |
| 05/03/2002 15:30 | 40.93 | 20.80 | 303 | 5 | 3.3 | 35.9 | 26/03/2002 19:11 | 35.35 | 31.28 | 1174 | 10 | 4.3 | 3.1 |
| 05/03/2002 18:30 | 40.76 | 20.78 | 315 | 10 | 3.5 | 35.4 | 26/03/2002 19:17 | 38.32 | 26.81 | 587 | 40 | 3.6 | 10.5 |
| 05/03/2002 21:03 | 38.74 | 22.76 | 387 | 5 | 3.7 | 24.7 | 26/03/2002 20:47 | 40.07 | 20.54 | 380 | 8 | 3.2 | 22.2 |
| 06/03/2002 02:52 | 37.09 | 28.13 | 787 | 21 | 3.9 | 6.3 | 27/03/2002 04:32 | 38.59 | 23.82 | 406 | 19 | 2.8 | 17.0 |
| 06/03/2002 07:04 | 40.97 | 20.75 | 306 | 5 | 3.0 | 32.0 | 29/03/2002 05:33 | 39.77 | 20.72 | 387 | 9 | 3.2 | 21.4 |
| 06/03/2002 15:45 | 41.05 | 20.74 | 303 | 5 | 3.0 | 32.6 | 30/03/2002 09:26 | 36.84 | 27 | 728 | 34 | 3.7 | 7.0 |
| 07/03/2002 00:54 | 37.21 | 21.34 | 592 | 5 | 2.9 | 8.3 | 30/03/2002 19:47 | 38.14 | 20.14 | 566 | 5 | 2.8 | 8.8 |
| 07/03/2002 06:23 | 38.14 | 22.02 | 470 | 5 | 3.3 | 14.9 | 30/03/2002 22:01 | 40.42 | 25.78 | 346 | 29 | 3.4 | 28.4 |
| 07/03/2002 19:33 | 37.82 | 21.22 | 535 | 8 | 3.4 | 11.9 | 30/03/2002 22:15 | 40.42 | 25.8 | 348 | 29 | 3.2 | 26.4 |
| 07/03/2002 20:30 | 40.98 | 20.77 | 304 | 5 | 3.8 | 41.2 | 31/03/2002 20:08 | 35.81 | 23.63 | 711 | 33 | 3.5 | 6.9 |
| 08/03/2002 13:31 | 41.01 | 23.38 | 133 | 8 | 3.6 | 202.7 | 31/03/2002 20:37 | 38.84 | 19.54 | 553 | 21 | 3.1 | 10.1 |

# April 2002 Report

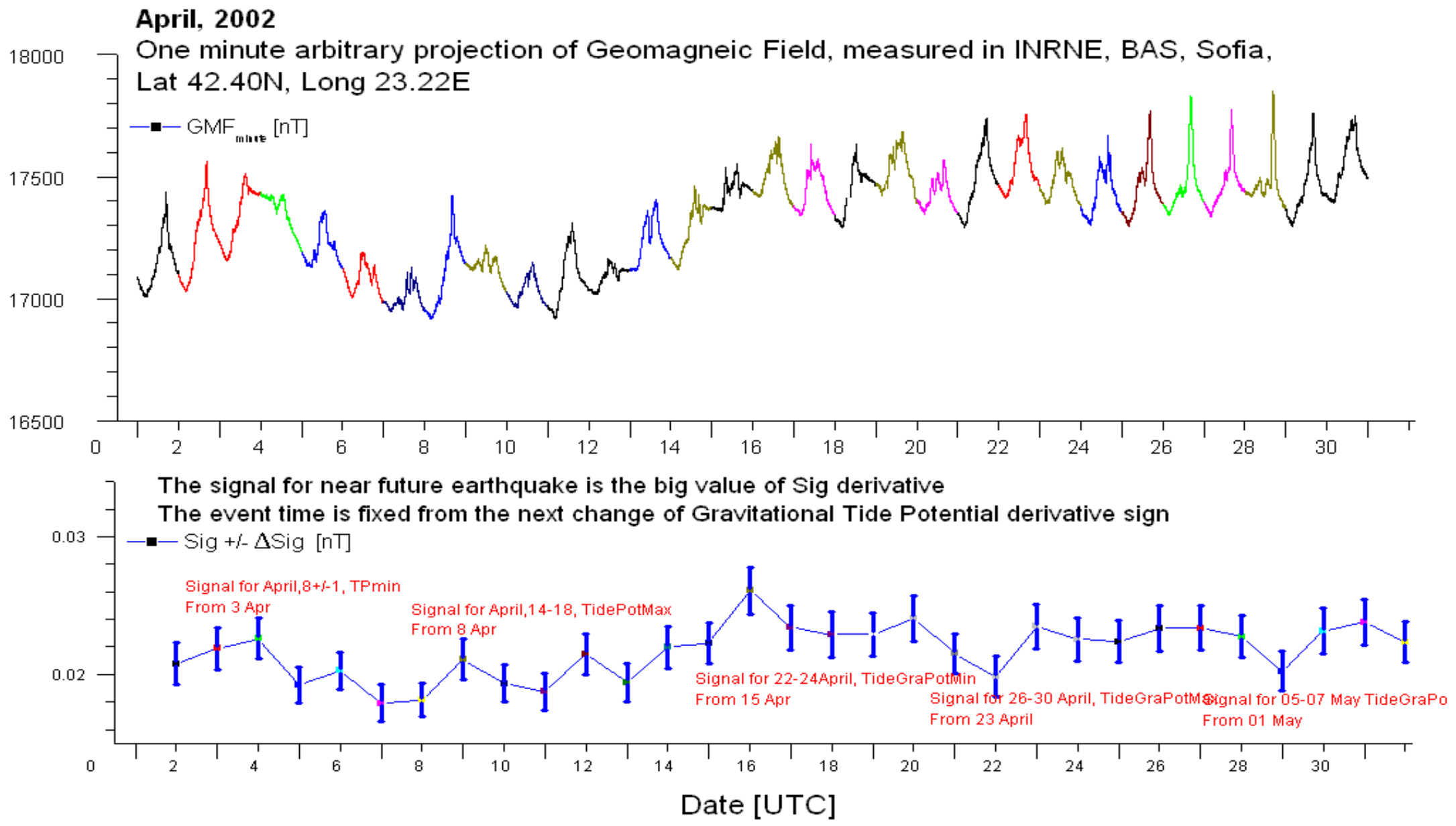

Figure April2002GMFSignal

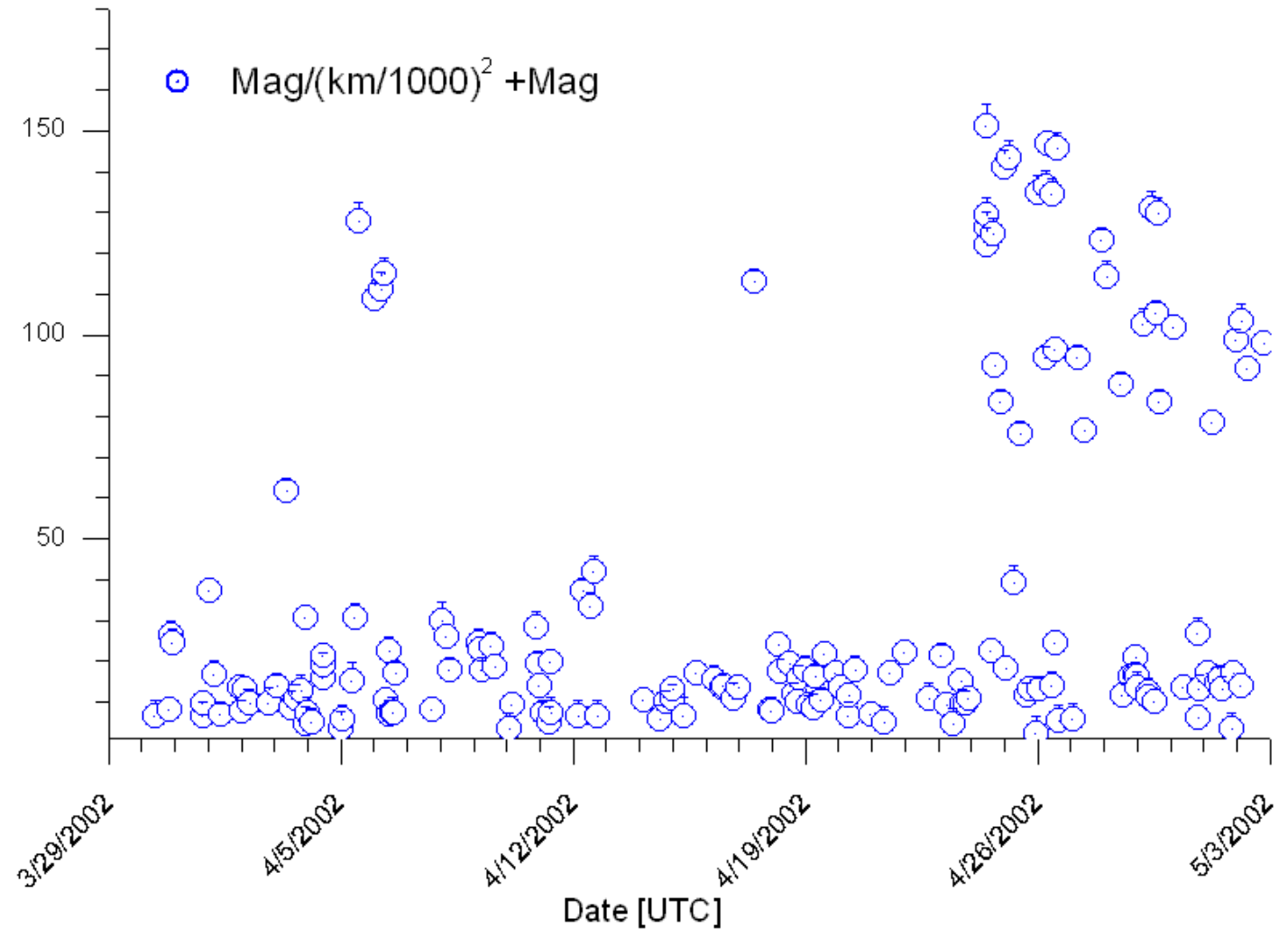

Figure April2002SChtM

| EQTable2002 | April | | | | | | | | | | | | |
|---|---|---|---|---|---|---|---|---|---|---|---|---|---|
| ddmmyy hhmm | Lat | Lon | kmSof | Dep | Mag | SChtM | ddmmyy hhmm | Lat | Lon | kmSof | Dep | Mag | SChtM |
| 01/04/2002 01:11 | 39.88 | 23.88 | 289 | 21 | 3.1 | 37.1 | 18/04/2002 19:38 | 38.50 | 23.39 | 433 | 10 | 3.0 | 16.0 |
| 01/04/2002 04:25 | 39.04 | 20.59 | 474 | 5 | 3.7 | 16.5 | 18/04/2002 23:52 | 38.62 | 23.64 | 422 | 10 | 3.2 | 18.0 |
| 01/04/2002 09:06 | 36.90 | 21.96 | 626 | 5 | 2.7 | 6.9 | 19/04/2002 02:59 | 37.73 | 22.02 | 535 | 3 | 2.6 | 9.1 |
| 01/04/2002 23:16 | 38.55 | 21.50 | 468 | 5 | 3.0 | 13.7 | 19/04/2002 05:46 | 36.84 | 21.55 | 644 | 38 | 3.5 | 8.4 |

| | | | | | | |
|---|---|---|---|---|---|---|
| 02/04/2002 00:46 | 37.01 | 23.10 | 598 | 5 | 2.8 | 7.8 |
| 02/04/2002 02:19 | 38.51 | 21.62 | 467 | 5 | 2.9 | 13.3 |
| 02/04/2002 06:26 | 37.04 | 22.30 | 604 | 5 | 3.5 | 9.6 |
| 02/04/2002 19:45 | 39.60 | 28.26 | 640 | 10 | 3.9 | 9.5 |
| 03/04/2002 02:01 | 39.21 | 20.71 | 451 | 5 | 2.8 | 13.8 |
| 03/04/2002 09:14 | 40.60 | 23.89 | 213 | 12 | 2.8 | 61.6 |
| 03/04/2002 12:00 | 35.80 | 23.71 | 735 | 5 | 4.5 | 8.3 |
| 03/04/2002 15:43 | 37.84 | 21.07 | 560 | 5 | 3.4 | 10.9 |
| 03/04/2002 19:26 | 38.20 | 20.70 | 544 | 5 | 3.8 | 12.9 |
| 03/04/2002 22:35 | 40.96 | 20.80 | 313 | 5 | 3.0 | 30.7 |
| 03/04/2002 22:57 | 37.76 | 30.51 | 959 | 49 | 4.3 | 4.7 |
| 04/04/2002 00:22 | 36.68 | 25.78 | 696 | 8 | 3.6 | 7.4 |
| 04/04/2002 03:05 | 35.37 | 26.25 | 850 | 22 | 3.6 | 5.0 |
| 04/04/2002 11:20 | 39.66 | 20.51 | 428 | 4 | 2.9 | 15.9 |
| 04/04/2002 11:24 | 38.90 | 22.16 | 406 | 5 | 3.1 | 18.8 |
| 04/04/2002 11:27 | 39.00 | 22.21 | 394 | 2 | 3.3 | 21.3 |
| 05/04/2002 00:38 | 40.62 | 33.05 | 1109 | 10 | 4.2 | 3.4 |
| 05/04/2002 01:27 | 35.35 | 24.26 | 791 | 38 | 3.5 | 5.6 |
| 05/04/2002 07:55 | 37.97 | 21.09 | 546 | 5 | 4.5 | 15.1 |
| 05/04/2002 10:27 | 41.90 | 20.36 | 322 | 10 | 3.2 | 30.8 |
| 05/04/2002 13:14 | 42.04 | 24.91 | 192 | 26 | 4.7 | 127.8 |
| 06/04/2002 01:10 | 42.06 | 24.87 | 187 | 10 | 3.8 | 108.7 |
| 06/04/2002 05:47 | 42.09 | 24.90 | 190 | 10 | 4.0 | 111.2 |
| 06/04/2002 07:48 | 42.10 | 24.83 | 182 | 10 | 3.8 | 115.0 |
| 06/04/2002 08:27 | 37.50 | 22.07 | 559 | 5 | 3.2 | 10.3 |
| 06/04/2002 11:19 | 35.78 | 23.88 | 739 | 5 | 3.7 | 6.8 |
| 06/04/2002 11:31 | 39.03 | 23.59 | 376 | 5 | 3.2 | 22.6 |
| 06/04/2002 13:49 | 36.07 | 21.55 | 727 | 5 | 3.9 | 7.4 |
| 06/04/2002 14:09 | 35.98 | 21.46 | 739 | 5 | 3.9 | 7.1 |
| 06/04/2002 15:54 | 39.67 | 20.78 | 406 | 25 | 2.8 | 17.0 |
| 07/04/2002 18:18 | 37.31 | 20.62 | 634 | 5 | 3.2 | 8.0 |
| 08/04/2002 01:20 | 42.43 | 19.77 | 383 | 10 | 4.4 | 30.0 |
| 08/04/2002 04:40 | 40.33 | 20.97 | 339 | 5 | 3.0 | 26.1 |
| 08/04/2002 06:37 | 42.43 | 19.78 | 382 | 8 | 2.6 | 17.8 |
| 09/04/2002 04:26 | 42.48 | 19.75 | 385 | 10 | 3.6 | 24.3 |
| 09/04/2002 04:39 | 42.46 | 19.79 | 381 | 10 | 3.3 | 22.8 |
| 09/04/2002 05:47 | 42.43 | 19.76 | 384 | 10 | 2.6 | 17.6 |
| 09/04/2002 13:02 | 42.39 | 19.80 | 380 | 8 | 3.4 | 23.6 |
| 09/04/2002 15:53 | 39.00 | 24.21 | 393 | 31 | 2.9 | 18.8 |
| 10/04/2002 02:20 | 36.31 | 30.89 | 1087 | 64 | 3.9 | 3.3 |
| 10/04/2002 03:59 | 37.28 | 21.19 | 611 | 5 | 3.4 | 9.1 |
| 10/04/2002 21:15 | 42.39 | 19.79 | 381 | 6 | 4.1 | 28.3 |
| 10/04/2002 23:07 | 42.41 | 19.78 | 382 | 0 | 2.8 | 19.2 |
| 10/04/2002 23:39 | 40.38 | 27.13 | 489 | 10 | 3.3 | 13.8 |
| 11/04/2002 02:50 | 37.12 | 21.00 | 636 | 5 | 2.9 | 7.2 |
| 11/04/2002 07:01 | 34.64 | 24.75 | 878 | 5 | 3.8 | 4.9 |
| 11/04/2002 08:03 | 38.88 | 24.52 | 417 | 5 | 3.4 | 19.6 |
| 11/04/2002 08:23 | 36.01 | 22.05 | 721 | 5 | 3.8 | 7.3 |
| 12/04/2002 03:54 | 35.79 | 22.80 | 735 | 5 | 3.6 | 6.7 |
| 12/04/2002 07:40 | 42.08 | 20.77 | 274 | 16 | 2.8 | 37.2 |
| 12/04/2002 12:40 | 40.94 | 20.80 | 314 | 10 | 3.3 | 33.5 |
| 12/04/2002 14:55 | 40.25 | 24.99 | 309 | 30 | 4.0 | 41.9 |
| 12/04/2002 17:55 | 35.51 | 23.59 | 766 | 5 | 3.8 | 6.5 |
| 14/04/2002 03:09 | 37.95 | 21.08 | 548 | 5 | 3.1 | 10.3 |
| 14/04/2002 15:00 | 36.16 | 26.93 | 806 | 5 | 3.7 | 5.7 |
| 14/04/2002 19:25 | 38.33 | 20.60 | 537 | 5 | 2.9 | 10.1 |
| 15/04/2002 00:14 | 38.43 | 26.72 | 588 | 24 | 3.7 | 10.7 |
| 15/04/2002 00:14 | 38.37 | 21.97 | 468 | 5 | 2.9 | 13.2 |
| 15/04/2002 08:10 | 34.83 | 24.59 | 854 | 56 | 4.7 | 6.5 |
| 15/04/2002 17:05 | 38.60 | 23.78 | 426 | 3 | 3.1 | 17.1 |
| 16/04/2002 06:46 | 40.29 | 19.73 | 453 | 10 | 3.2 | 15.6 |
| 16/04/2002 12:35 | 38.98 | 20.63 | 476 | 5 | 3.1 | 13.7 |
| 16/04/2002 12:51 | 38.95 | 20.71 | 474 | 22 | 3.1 | 13.8 |
| 16/04/2002 19:59 | 38.01 | 20.23 | 590 | 57 | 3.8 | 10.9 |
| 16/04/2002 23:41 | 38.81 | 26.01 | 505 | 29 | 3.4 | 13.4 |
| 17/04/2002 11:18 | 41.66 | 24.43 | 157 | 10 | 2.8 | 113.0 |
| 17/04/2002 22:27 | 37.21 | 21.29 | 615 | 13 | 3.0 | 7.9 |
| 17/04/2002 23:50 | 37.86 | 27.28 | 676 | 11 | 3.5 | 7.7 |
| 18/04/2002 04:55 | 39.40 | 21.87 | 365 | 7 | 3.2 | 24.0 |
| 18/04/2002 06:15 | 39.50 | 20.64 | 431 | 13 | 3.2 | 17.2 |
| 18/04/2002 13:11 | 38.98 | 22.14 | 398 | 16 | 3.1 | 19.6 |
| 18/04/2002 16:01 | 38.05 | 22.93 | 484 | 30 | 2.8 | 12.0 |
| 18/04/2002 18:21 | 38.11 | 20.81 | 546 | 5 | 3.0 | 10.1 |
| 19/04/2002 07:30 | 38.53 | 23.53 | 431 | 23 | 3.0 | 16.2 |
| 19/04/2002 11:52 | 38.30 | 20.61 | 540 | 5 | 3.0 | 10.3 |
| 19/04/2002 13:45 | 39.07 | 24.14 | 384 | 58 | 3.2 | 21.8 |
| 19/04/2002 21:51 | 38.57 | 23.71 | 429 | 25 | 3.1 | 16.9 |
| 20/04/2002 01:21 | 38.53 | 25.46 | 496 | 40 | 3.3 | 13.4 |
| 20/04/2002 07:33 | 35.93 | 22.09 | 729 | 27 | 3.5 | 6.6 |
| 20/04/2002 07:35 | 38.06 | 21.82 | 506 | 22 | 3.0 | 11.7 |
| 20/04/2002 12:27 | 38.62 | 23.62 | 422 | 10 | 3.2 | 18.0 |
| 20/04/2002 23:34 | 36.13 | 23.85 | 700 | 5 | 3.3 | 6.7 |
| 21/04/2002 09:20 | 34.97 | 26.55 | 904 | 68 | 4.0 | 4.9 |
| 21/04/2002 13:09 | 38.57 | 23.53 | 427 | 22 | 3.1 | 17.0 |
| 21/04/2002 23:34 | 39.29 | 21.94 | 373 | 6 | 3.1 | 22.3 |
| 22/04/2002 17:54 | 39.25 | 27.58 | 597 | 10 | 3.9 | 10.9 |
| 23/04/2002 02:18 | 41.26 | 20.15 | 364 | 9 | 2.8 | 21.2 |
| 23/04/2002 06:16 | 37.73 | 21.00 | 574 | 5 | 3.1 | 9.4 |
| 23/04/2002 11:10 | 36.23 | 29.01 | 939 | 57 | 3.9 | 4.4 |
| 23/04/2002 17:12 | 39.24 | 20.63 | 454 | 10 | 3.1 | 15.1 |
| 23/04/2002 18:02 | 37.38 | 21.26 | 598 | 5 | 3.5 | 9.8 |
| 23/04/2002 20:13 | 37.41 | 21.47 | 587 | 5 | 3.3 | 9.6 |
| 23/04/2002 22:10 | 37.71 | 21.69 | 548 | 10 | 3.2 | 10.7 |
| 24/04/2002 10:51 | 42.44 | 21.47 | 194 | 10 | 5.7 | 151.0 |
| 24/04/2002 11:17 | 42.46 | 21.68 | 171 | 10 | 3.7 | 126.4 |
| 24/04/2002 11:24 | 42.44 | 21.53 | 188 | 10 | 4.3 | 122.1 |
| 24/04/2002 11:33 | 42.51 | 21.62 | 178 | 10 | 4.1 | 129.4 |
| 24/04/2002 14:21 | 41.10 | 20.26 | 359 | 10 | 2.9 | 22.5 |
| 24/04/2002 16:04 | 42.50 | 21.63 | 177 | 10 | 3.9 | 124.7 |
| 24/04/2002 17:09 | 42.38 | 21.57 | 183 | 10 | 3.1 | 92.4 |
| 24/04/2002 21:12 | 42.47 | 21.57 | 183 | 10 | 2.8 | 83.3 |
| 24/04/2002 23:37 | 42.51 | 21.65 | 175 | 10 | 4.3 | 140.9 |
| 25/04/2002 01:12 | 39.76 | 25.85 | 414 | 39 | 3.1 | 18.1 |
| 25/04/2002 03:43 | 42.49 | 21.70 | 169 | 10 | 4.1 | 143.5 |
| 25/04/2002 07:28 | 40.67 | 20.83 | 328 | 10 | 4.2 | 39.2 |
| 25/04/2002 11:56 | 42.37 | 21.52 | 189 | 10 | 2.7 | 75.8 |
| 25/04/2002 16:30 | 38.13 | 21.64 | 505 | 4 | 3.0 | 11.8 |
| 25/04/2002 18:01 | 38.99 | 20.75 | 467 | 27 | 2.9 | 13.3 |
| 25/04/2002 22:34 | 35.19 | 32.63 | 1316 | 78 | 4.2 | 2.4 |
| 26/04/2002 00:21 | 42.46 | 21.65 | 174 | 10 | 4.1 | 134.8 |
| 26/04/2002 01:28 | 38.44 | 21.70 | 471 | 19 | 2.9 | 13.1 |
| 26/04/2002 05:57 | 42.40 | 21.64 | 175 | 10 | 2.9 | 94.3 |
| 26/04/2002 06:31 | 42.49 | 21.68 | 171 | 10 | 4.0 | 136.4 |
| 26/04/2002 07:00 | 42.46 | 21.87 | 150 | 10 | 3.3 | 146.7 |
| 26/04/2002 10:10 | 39.48 | 19.66 | 511 | 12 | 3.6 | 13.8 |
| 26/04/2002 10:17 | 42.50 | 21.75 | 164 | 10 | 3.6 | 134.6 |
| 26/04/2002 12:37 | 41.00 | 20.32 | 357 | 10 | 3.1 | 24.3 |
| 26/04/2002 12:42 | 42.44 | 21.63 | 177 | 10 | 3.0 | 96.3 |
| 26/04/2002 14:14 | 42.51 | 21.77 | 161 | 10 | 3.8 | 145.9 |
| 26/04/2002 15:28 | 35.90 | 27.36 | 855 | 23 | 3.9 | 5.3 |
| 27/04/2002 01:41 | 34.85 | 24.27 | 846 | 56 | 4.0 | 5.6 |
| 27/04/2002 05:04 | 42.41 | 21.64 | 175 | 10 | 2.9 | 94.3 |
| 27/04/2002 10:16 | 42.33 | 21.50 | 191 | 10 | 2.8 | 76.7 |
| 27/04/2002 22:23 | 42.65 | 21.91 | 148 | 10 | 2.7 | 123.2 |
| 28/04/2002 02:30 | 42.43 | 21.60 | 180 | 10 | 3.7 | 114.4 |
| 28/04/2002 12:09 | 42.40 | 21.61 | 179 | 10 | 2.8 | 87.7 |
| 28/04/2002 13:13 | 38.64 | 20.45 | 518 | 7 | 3.1 | 11.5 |
| 28/04/2002 19:15 | 38.60 | 23.90 | 429 | 19 | 3.0 | 16.3 |
| 28/04/2002 22:41 | 38.54 | 23.62 | 431 | 30 | 3.1 | 16.7 |
| 28/04/2002 22:59 | 39.95 | 20.67 | 393 | 10 | 3.2 | 20.8 |
| 29/04/2002 00:17 | 38.80 | 21.84 | 428 | 5 | 3.0 | 16.4 |
| 29/04/2002 00:24 | 38.51 | 20.51 | 526 | 9 | 3.8 | 13.7 |
| 29/04/2002 04:39 | 42.44 | 21.51 | 190 | 10 | 3.7 | 102.6 |
| 29/04/2002 08:08 | 38.52 | 20.44 | 530 | 5 | 3.5 | 12.5 |
| 29/04/2002 08:24 | 38.50 | 20.43 | 532 | 16 | 3.2 | 11.3 |
| 29/04/2002 10:10 | 42.50 | 21.61 | 179 | 10 | 4.2 | 131.0 |
| 29/04/2002 12:45 | 38.34 | 20.45 | 546 | 36 | 3.0 | 10.1 |
| 29/04/2002 13:58 | 42.48 | 21.70 | 169 | 10 | 3.0 | 105.1 |
| 29/04/2002 15:37 | 42.48 | 21.66 | 173 | 10 | 3.9 | 129.7 |
| 29/04/2002 15:53 | 42.41 | 21.60 | 180 | 10 | 2.7 | 83.5 |
| 30/04/2002 02:23 | 42.49 | 21.65 | 175 | 10 | 3.1 | 101.7 |
| 30/04/2002 09:36 | 38.05 | 23.95 | 490 | 11 | 3.2 | 13.4 |
| 30/04/2002 19:46 | 39.21 | 21.83 | 386 | 5 | 4.0 | 26.8 |
| 30/04/2002 20:13 | 35.87 | 23.52 | 726 | 29 | 3.2 | 6.1 |
| 30/04/2002 20:57 | 38.45 | 25.38 | 500 | 23 | 3.3 | 13.2 |

# May 2002 Report

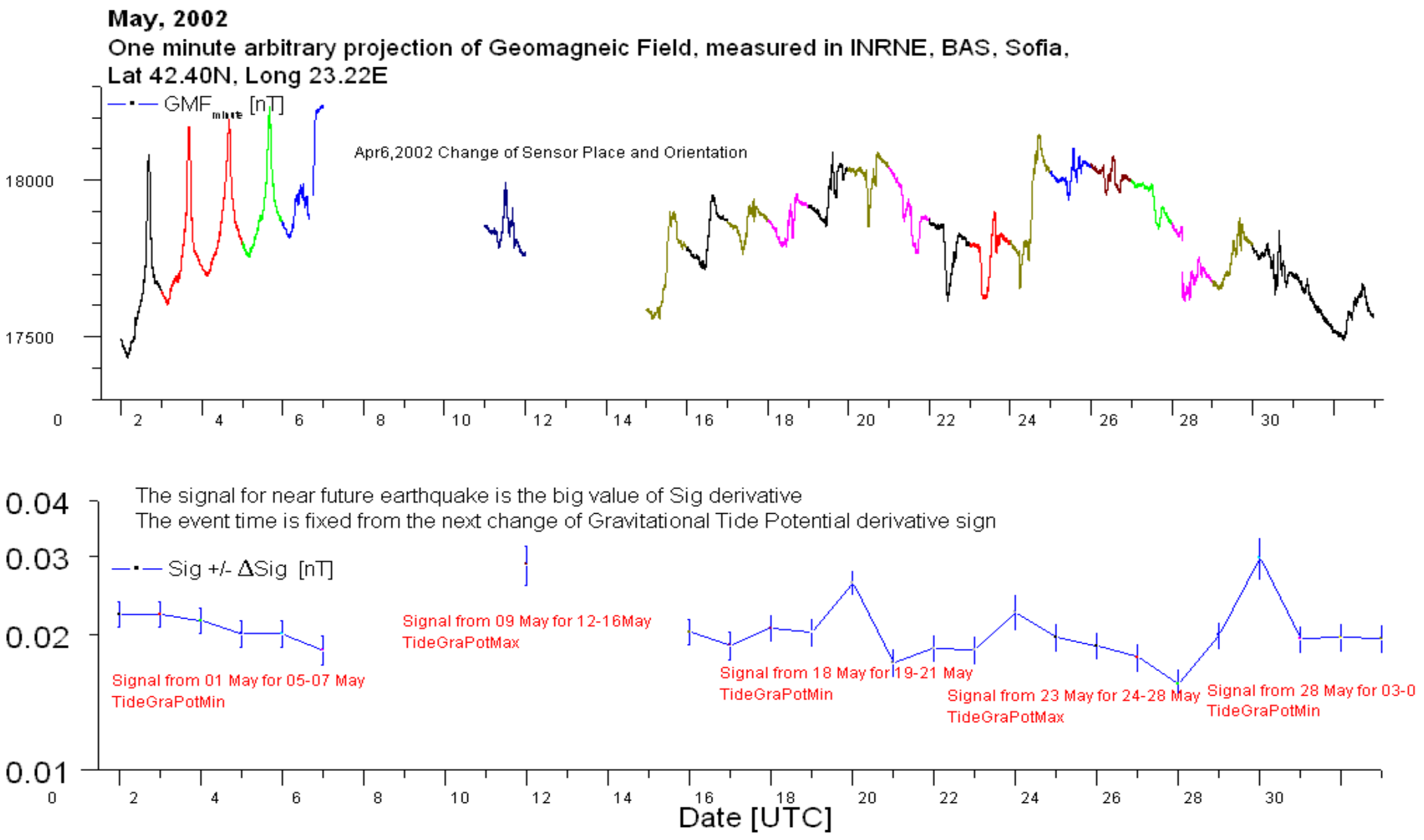

Figure May2002GMFSignal

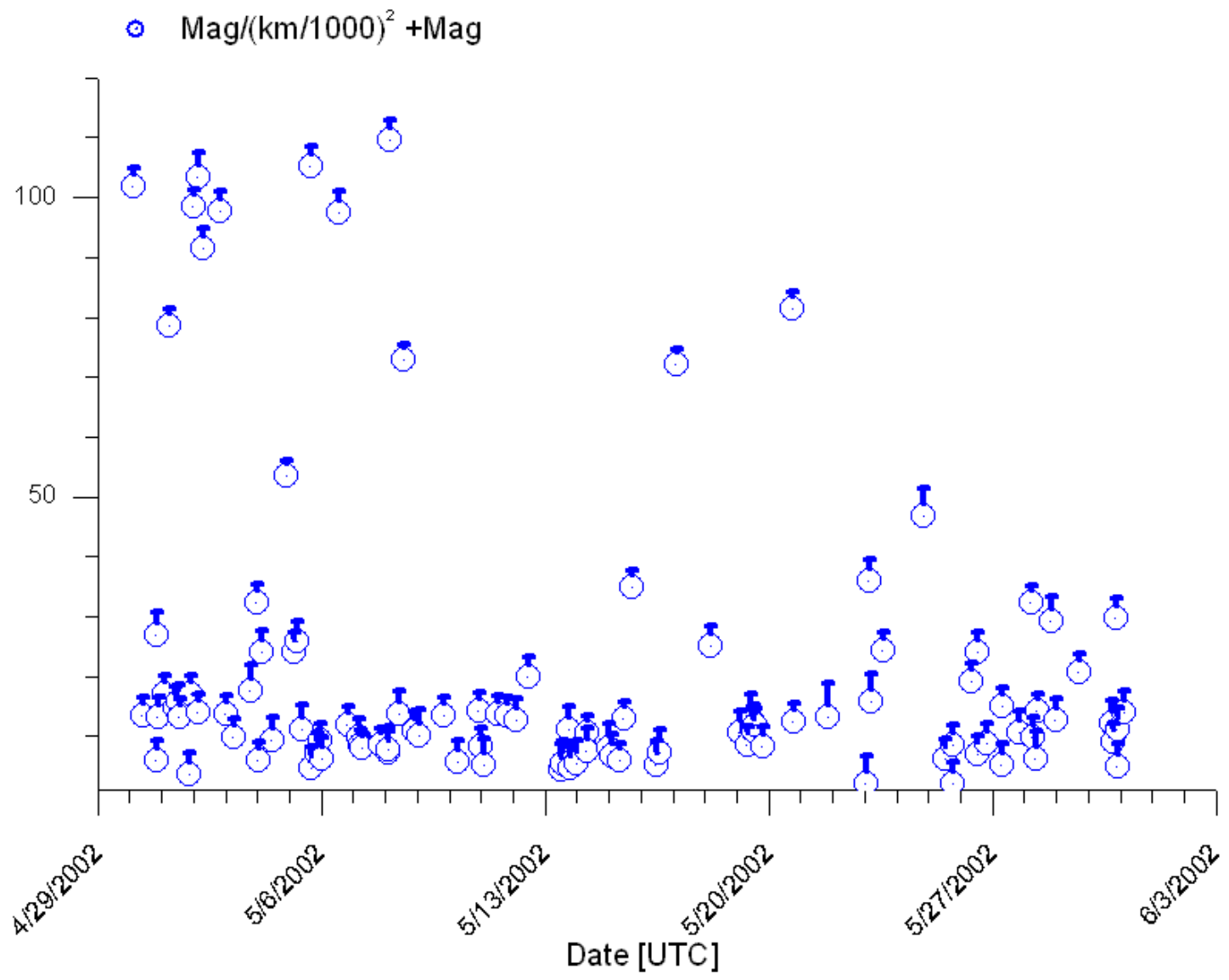

Figure May2002SChtM

| EQtable2002 | May | | | | | | | | | | | | |
|---|---|---|---|---|---|---|---|---|---|---|---|---|---|
| ddmmyy hhmm | Lat | Lon | kmSofia | Dep | Mag | SChtM | ddmmyy hhmm | Lat | Lon | kmSofia | Dep | Mag | SChtM |
| 01/05/2002 02:35 | 38.50 | 23.40 | 433 | 5 | 3.2 | 17.0 | 13/05/2002 18:30 | 35.35 | 26.80 | 878 | 17 | 3.6 | 4.7 |
| 01/05/2002 06:03 | 42.45 | 21.52 | 189 | 10 | 2.8 | 78.6 | 13/05/2002 23:39 | 35.66 | 26.64 | 839 | 10 | 3.9 | 5.5 |
| 01/05/2002 10:30 | 39.54 | 19.72 | 502 | 32 | 3.7 | 14.7 | 14/05/2002 07:15 | 37.97 | 21.56 | 525 | 3 | 2.9 | 10.5 |
| 01/05/2002 12:56 | 38.57 | 23.66 | 428 | 20 | 2.9 | 15.8 | 14/05/2002 07:28 | 36.71 | 26.33 | 720 | 23 | 3.9 | 7.5 |
| 01/05/2002 13:40 | 38.12 | 23.46 | 476 | 5 | 3.0 | 13.3 | 14/05/2002 23:32 | 36.96 | 26.62 | 712 | 33 | 4.1 | 8.1 |
| 01/05/2002 20:51 | 36.45 | 30.42 | 1037 | 76 | 3.8 | 3.5 | 15/05/2002 02:28 | 36.78 | 26.59 | 727 | 5 | 3.6 | 6.8 |
| 01/05/2002 22:06 | 38.65 | 23.67 | 419 | 5 | 3.0 | 17.1 | 15/05/2002 08:14 | 36.27 | 21.80 | 698 | 5 | 2.9 | 5.9 |
| 01/05/2002 23:46 | 42.45 | 21.73 | 166 | 10 | 2.7 | 98.6 | 15/05/2002 11:18 | 39.31 | 20.30 | 472 | 31 | 2.9 | 13.0 |
| 02/05/2002 03:10 | 38.42 | 22.01 | 462 | 19 | 3.0 | 14.1 | 15/05/2002 17:24 | 39.88 | 24.02 | 294 | 5 | 3.0 | 34.8 |
| 02/05/2002 03:31 | 42.47 | 21.45 | 197 | 10 | 4.0 | 103.5 | 16/05/2002 11:19 | 35.65 | 27.49 | 887 | 10 | 4.1 | 5.2 |
| 02/05/2002 07:32 | 42.40 | 21.51 | 190 | 10 | 3.3 | 91.6 | 16/05/2002 14:11 | 36.92 | 26.64 | 717 | 5 | 3.8 | 7.4 |
| 02/05/2002 20:13 | 42.40 | 21.59 | 181 | 10 | 3.2 | 97.8 | 17/05/2002 02:45 | 42.39 | 21.51 | 190 | 10 | 2.6 | 72.2 |
| 03/05/2002 00:12 | 38.39 | 21.99 | 466 | 17 | 3.0 | 13.8 | 18/05/2002 04:44 | 40.83 | 20.42 | 356 | 10 | 3.2 | 25.2 |
| 03/05/2002 06:01 | 37.79 | 21.09 | 564 | 4 | 3.1 | 9.8 | 19/05/2002 02:12 | 39.20 | 27.59 | 601 | 5 | 3.8 | 10.5 |
| 03/05/2002 18:31 | 45.69 | 26.29 | 500 | 152 | 4.4 | 17.6 | 19/05/2002 07:35 | 37.26 | 21.95 | 588 | 5 | 3.0 | 8.7 |
| 03/05/2002 23:03 | 43.20 | 20.55 | 309 | 8 | 3.1 | 32.4 | 19/05/2002 10:45 | 38.38 | 26.39 | 568 | 28 | 4.2 | 13.0 |
| 04/05/2002 00:23 | 35.93 | 22.16 | 728 | 13 | 3.1 | 5.9 | 19/05/2002 11:07 | 38.39 | 26.59 | 581 | 36 | 3.8 | 11.2 |
| 04/05/2002 03:29 | 39.18 | 21.76 | 392 | 13 | 3.7 | 24.0 | 19/05/2002 12:21 | 38.40 | 26.50 | 574 | 37 | 3.8 | 11.5 |
| 04/05/2002 11:02 | 39.12 | 27.93 | 637 | 10 | 3.8 | 9.4 | 19/05/2002 12:26 | 38.35 | 26.30 | 565 | 21 | 3.5 | 11.0 |
| 04/05/2002 21:56 | 42.55 | 21.20 | 225 | 9 | 2.7 | 53.4 | 19/05/2002 12:49 | 38.36 | 26.33 | 566 | 21 | 3.5 | 10.9 |
| 05/05/2002 03:01 | 39.16 | 24.20 | 376 | 34 | 3.4 | 24.1 | 19/05/2002 13:56 | 38.14 | 22.26 | 485 | 28 | 2.8 | 11.9 |
| 05/05/2002 06:15 | 41.25 | 20.22 | 357 | 0 | 3.3 | 26.0 | 19/05/2002 19:04 | 37.71 | 26.92 | 663 | 29 | 3.6 | 8.2 |
| 05/05/2002 09:22 | 40.55 | 28.33 | 603 | 10 | 4.1 | 11.3 | 20/05/2002 18:09 | 42.37 | 21.58 | 182 | 10 | 2.7 | 81.5 |
| 05/05/2002 16:06 | 42.39 | 21.60 | 180 | 10 | 3.4 | 105.1 | 20/05/2002 18:24 | 38.41 | 25.45 | 507 | 16 | 3.2 | 12.4 |
| 05/05/2002 16:16 | 34.79 | 26.05 | 901 | 18 | 3.7 | 4.6 | 21/05/2002 20:53 | 36.63 | 24.27 | 651 | 97 | 5.6 | 13.2 |
| 05/05/2002 20:39 | 36.52 | 25.59 | 704 | 24 | 3.5 | 7.1 | 23/05/2002 01:08 | 37.39 | 36.34 | 1559 | 8 | 4.8 | 2.0 |
| 05/05/2002 22:41 | 37.84 | 21.04 | 561 | 4 | 2.9 | 9.2 | 23/05/2002 03:25 | 44.73 | 21.65 | 312 | 10 | 3.5 | 36.0 |
| 06/05/2002 00:28 | 35.61 | 24.01 | 759 | 41 | 3.6 | 6.3 | 23/05/2002 04:48 | 38.76 | 26.42 | 538 | 14 | 4.6 | 15.9 |
| 06/05/2002 13:29 | 42.35 | 21.49 | 192 | 10 | 3.6 | 97.5 | 23/05/2002 13:44 | 42.53 | 26.32 | 344 | 10 | 2.9 | 24.5 |
| 06/05/2002 19:27 | 37.96 | 22.05 | 510 | 28 | 3.1 | 11.9 | 24/05/2002 20:42 | 44.76 | 21.61 | 317 | 10 | 4.7 | 46.7 |
| 07/05/2002 04:30 | 37.88 | 21.21 | 549 | 29 | 3.0 | 10.0 | 25/05/2002 12:31 | 35.90 | 23.18 | 722 | 5 | 3.3 | 6.3 |
| 07/05/2002 05:27 | 37.88 | 21.18 | 551 | 5 | 2.6 | 8.6 | 25/05/2002 18:30 | 37.10 | 21.56 | 617 | 9 | 3.3 | 8.7 |
| 07/05/2002 06:58 | 37.41 | 22.01 | 570 | 16 | 2.6 | 8.0 | 25/05/2002 18:43 | 34.92 | 32.98 | 1365 | 57 | 3.8 | 2.0 |
| 07/05/2002 19:47 | 37.38 | 22.05 | 572 | 20 | 2.8 | 8.6 | 26/05/2002 08:13 | 40.07 | 20.53 | 395 | 10 | 3.0 | 19.2 |
| 08/05/2002 02:00 | 36.91 | 26.65 | 719 | 30 | 3.7 | 7.2 | 26/05/2002 12:51 | 39.19 | 24.45 | 382 | 16 | 3.5 | 24.0 |
| 08/05/2002 02:29 | 36.91 | 26.63 | 717 | 31 | 4.0 | 7.8 | 26/05/2002 12:54 | 36.17 | 23.52 | 692 | 5 | 3.3 | 6.9 |
| 08/05/2002 03:45 | 42.41 | 21.61 | 179 | 10 | 3.5 | 109.6 | 26/05/2002 20:08 | 37.08 | 21.72 | 614 | 3 | 3.3 | 8.8 |
| 08/05/2002 10:23 | 38.59 | 25.90 | 517 | 15 | 3.7 | 13.8 | 27/05/2002 07:13 | 37.03 | 28.20 | 813 | 14 | 3.5 | 5.3 |
| 08/05/2002 13:58 | 42.33 | 21.52 | 189 | 10 | 2.6 | 72.9 | 27/05/2002 07:24 | 38.48 | 24.13 | 447 | 5 | 3.0 | 15.0 |
| 08/05/2002 22:12 | 38.59 | 26.25 | 540 | 5 | 3.2 | 11.0 | 27/05/2002 20:21 | 37.87 | 25.99 | 589 | 5 | 3.7 | 10.7 |
| 09/05/2002 01:49 | 36.47 | 23.33 | 658 | 10 | 4.4 | 10.2 | 28/05/2002 04:47 | 40.21 | 21.56 | 305 | 22 | 3.0 | 32.2 |
| 09/05/2002 20:15 | 38.27 | 22.22 | 472 | 43 | 3.0 | 13.5 | 28/05/2002 06:20 | 37.44 | 21.66 | 577 | 7 | 3.3 | 9.9 |
| 10/05/2002 06:39 | 35.36 | 23.70 | 783 | 44 | 3.5 | 5.7 | 28/05/2002 08:26 | 36.02 | 27.35 | 844 | 104 | 4.5 | 6.3 |
| 10/05/2002 22:21 | 38.76 | 21.29 | 457 | 10 | 3.0 | 14.3 | 28/05/2002 09:51 | 38.87 | 21.05 | 460 | 14 | 3.0 | 14.2 |
| 10/05/2002 23:46 | 36.95 | 22.56 | 609 | 17 | 3.1 | 8.4 | 28/05/2002 19:45 | 42.44 | 19.76 | 384 | 33 | 4.3 | 29.2 |
| 11/05/2002 02:26 | 35.83 | 28.37 | 927 | 65 | 4.4 | 5.1 | 28/05/2002 23:32 | 38.66 | 26.22 | 532 | 24 | 3.6 | 12.7 |
| 11/05/2002 12:43 | 39.86 | 19.88 | 466 | 18 | 3.0 | 13.8 | 29/05/2002 16:46 | 38.99 | 24.17 | 393 | 34 | 3.2 | 20.7 |
| 11/05/2002 18:51 | 38.66 | 21.38 | 463 | 31 | 2.9 | 13.6 | 30/05/2002 17:18 | 37.53 | 21.94 | 559 | 13 | 3.8 | 12.2 |
| 12/05/2002 02:31 | 38.57 | 20.30 | 535 | 18 | 3.6 | 12.6 | 30/05/2002 19:12 | 37.33 | 21.21 | 605 | 10 | 3.3 | 9.0 |
| 12/05/2002 11:25 | 38.92 | 24.37 | 407 | 39 | 3.3 | 19.9 | 30/05/2002 20:29 | 40.96 | 20.48 | 344 | 5 | 3.5 | 29.7 |
| 13/05/2002 11:42 | 38.60 | 31.26 | 987 | 10 | 4.4 | 4.5 | 30/05/2002 21:44 | 37.53 | 21.95 | 559 | 8 | 3.5 | 11.2 |
| 13/05/2002 12:32 | 35.61 | 26.76 | 850 | 9 | 3.9 | 5.4 | 30/05/2002 21:57 | 34.57 | 22.44 | 873 | 37 | 3.8 | 5.0 |
| 13/05/2002 17:50 | 37.58 | 20.95 | 591 | 33 | 3.9 | 11.2 | 31/05/2002 02:52 | 39.09 | 20.28 | 491 | 29 | 3.4 | 14.1 |

## June 2002 Report

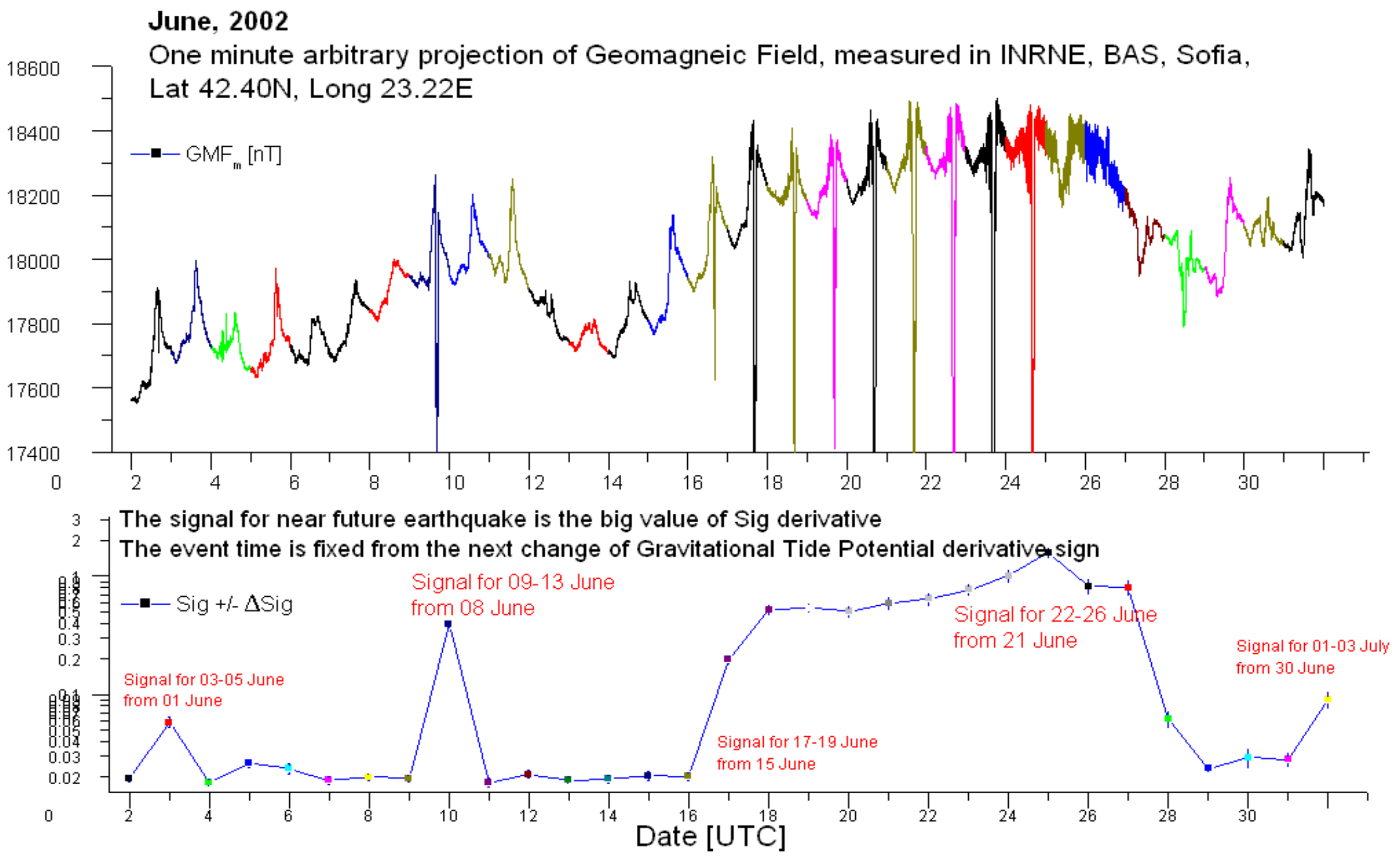

Figure June2002GMFSignal

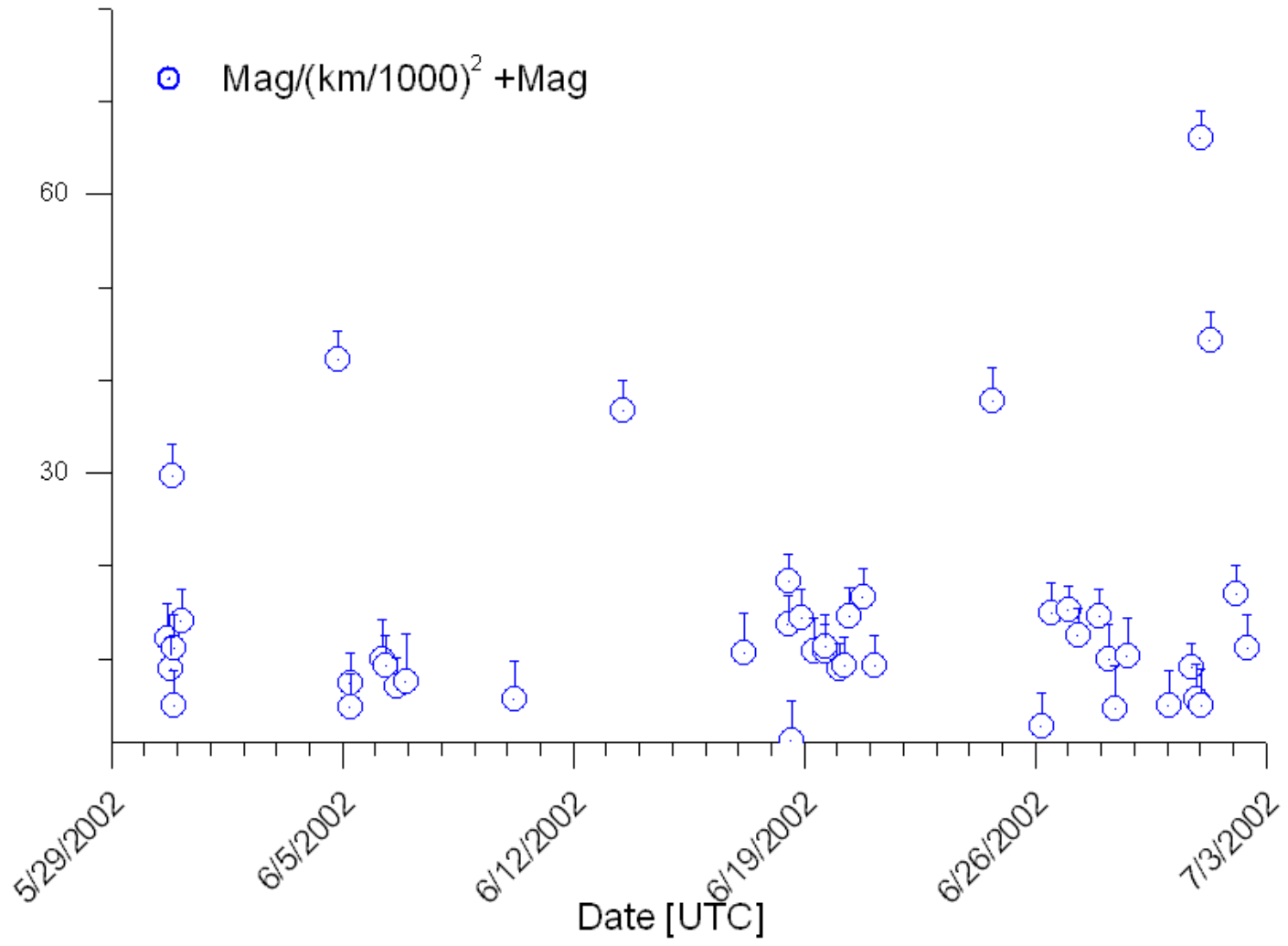

Figure June2002SChtM

**EQTable2002 June**

| ddmmyy hhmm | Lat | Lon | kmSof | Dep | Mag | SChtM | ddmmyy hhmm | Lat | Lon | kmSof | Dep | Mag | SChtM |
|---|---|---|---|---|---|---|---|---|---|---|---|---|---|
| 6/4/02 21:10 | 40.45 | 21.75 | 271 | 27 | 3.1 | 42.2 | 6/20/02 1:27 | 37.82 | 21.12 | 559 | 15 | 2.8 | 9.0 |
| 6/5/02 5:37 | 34.90 | 25.30 | 864 | 5 | 3.6 | 4.8 | 6/20/02 5:24 | 37.61 | 21.20 | 577 | 5 | 3.1 | 9.3 |
| 6/5/02 6:00 | 36.50 | 23.13 | 655 | 5 | 3.2 | 7.5 | 6/20/02 8:57 | 38.43 | 21.78 | 469 | 5 | 3.2 | 14.6 |
| 6/6/02 5:10 | 39.05 | 28.00 | 648 | 8 | 4.2 | 10.0 | 6/20/02 18:45 | 38.91 | 24.94 | 432 | 16 | 3.1 | 16.6 |
| 6/6/02 7:53 | 37.24 | 22.11 | 586 | 25 | 3.2 | 9.3 | 6/21/02 3:18 | 37.34 | 21.45 | 595 | 2 | 3.3 | 9.3 |
| 6/6/02 15:12 | 36.62 | 21.73 | 663 | 5 | 3.1 | 7.1 | 6/24/02 17:08 | 42.16 | 20.49 | 304 | 10 | 3.5 | 37.8 |
| 6/6/02 22:35 | 35.65 | 26.22 | 820 | 93 | 5.1 | 7.6 | 6/26/02 4:31 | 36.04 | 31.45 | 1155 | 10 | 3.6 | 2.7 |
| 6/10/02 5:20 | 34.65 | 23.33 | 860 | 63 | 4.2 | 5.7 | 6/26/02 12:18 | 38.42 | 21.74 | 471 | 5 | 3.3 | 14.9 |
| 6/13/02 12:43 | 40.50 | 21.30 | 300 | 10 | 3.3 | 36.7 | 6/27/02 0:27 | 38.70 | 23.63 | 413 | 5 | 2.6 | 15.2 |
| 6/17/02 4:43 | 36.68 | 22.22 | 645 | 33 | 4.4 | 10.6 | 6/27/02 7:29 | 38.15 | 24.08 | 481 | 18 | 2.9 | 12.5 |
| 6/18/02 12:38 | 38.33 | 21.87 | 476 | 5 | 3.1 | 13.7 | 6/27/02 22:41 | 38.48 | 22.82 | 437 | 4 | 2.8 | 14.6 |
| 6/18/02 13:10 | 39.12 | 21.77 | 398 | 5 | 2.9 | 18.3 | 6/28/02 5:36 | 36.95 | 22.18 | 616 | 8 | 3.8 | 10.0 |
| 6/18/02 14:58 | 39.34 | 39.46 | 1834 | 10 | 4.2 | 1.3 | 6/28/02 10:47 | 38.44 | 31.32 | 1001 | 33 | 4.6 | 4.6 |
| 6/18/02 22:21 | 38.37 | 22.10 | 464 | 5 | 3.1 | 14.4 | 6/28/02 19:23 | 36.73 | 23.22 | 629 | 79 | 4.1 | 10.4 |
| 6/19/02 7:14 | 37.70 | 20.97 | 578 | 5 | 3.6 | 10.8 | 6/30/02 1:14 | 35.23 | 26.52 | 876 | 5 | 3.8 | 5.0 |
| 6/19/02 15:29 | 38.20 | 21.20 | 517 | 13 | 2.9 | 10.8 | 6/30/02 18:02 | 37.57 | 22.26 | 547 | 18 | 2.7 | 9.0 |
| 6/19/02 15:29 | 38.34 | 20.57 | 538 | 5 | 3.3 | 11.4 | 6/30/02 20:51 | 36.21 | 27.45 | 832 | 8 | 3.9 | 5.6 |